\documentclass[12pt]{JHEP3}
\usepackage{amsmath,amssymb,cite,scalefnt}
\usepackage[dvips]{epsfig}
\def\besub{\begin{subequations}}
\def\eesub{\end{subequations}}
\def\be{\begin{equation}}
\def\ee{\end{equation}}
\def\bea{\begin{eqnarray}}
\def\eea{\end{eqnarray}}
\def\frak#1#2{{\textstyle{\frac{#1}{#2}}}}
\def\Ycal{{\cal Y}}
\def\lf{16\pi^2}
\def\Tr{{\rm Tr }}

\newcommand{\lae}{\lower 2pt \hbox{$\, \buildrel {\scriptstyle <}
\over {\scriptstyle
\sim}\,$}}
\newcommand{\gae}{\lower 2pt \hbox{$\, \buildrel {\scriptstyle >}
\over {\scriptstyle
\sim}\,$}}
\newcommand{\eqn}[1]{Eq.~(\ref{#1})}

\newcommand{\tabby}[1]{Table~\ref{#1}}
\newcommand{\reference}[1]{Ref.~\cite{#1}}
\newcommand{\abbrev}{\scalefont{.9}}
\newcommand{\susy}{{\abbrev SUSY}}
\newcommand{\cp}{{\abbrev CP}}
\newcommand{\ckm}{{\abbrev CKM}}
\newcommand{\sm}{{\abbrev SM}}

\newcommand{\mssm}{{\abbrev MSSM}}
\newcommand{\amsb}{{\abbrev AMSB}}
\newcommand{\mamsb}{{\abbrev mAMSB}}
\newcommand{\msugra}{{\abbrev mSUGRA}}

\newcommand{\mfv}{{\abbrev MFV}}

\newcommand{\RG}{{\abbrev RG}}
\newcommand{\fcnc}{{\abbrev FCNC}}
\newcommand{\irqfp}{{\abbrev IRQFP}}
\newcommand{\mbf}{\mathbf}
\def\fivebar{{\overline{5}}}
\def\lambdabar{{\overline{\lambda}}}
\def\sy{supersymmetry}
\def\sic{supersymmetric}
\def\mbar{{\overline{m}}}

\def\TeV{\hbox{TeV}}
\newcommand{\msbar}{\rm {\overline{MS}}}
\newcommand{\drbar}{\rm {\overline{DR}}}
\newcommand{\msoft}{m_{3/2}}
\newcommand{\bsg}{b \rightarrow s\gamma}
\newcommand{\Bsg}{B \rightarrow X_s\gamma}
\newcommand{\bratio}{{\rm BR}(\Bsg)}
\newcommand{\msusy}{M_{{\scriptscriptstyle \rm SUSY}}}
\newcommand{\mgut}{M_{{\scriptscriptstyle \rm GUT}}}
\newcommand{\mgl}{m_{\tilde{g}}}
\newcommand{\dq}{\delta^q}

\author{B C Allanach$^{1}$, G Hiller$^2$, D R T Jones$^{3,4}$, P Slavich$^{5,6}$\\
  $^{1}$ DAMTP, CMS, University of Cambridge, Wilberforce Road,
  Cambridge CB3
  0WA, United Kingdom\\
  $^{2}$ Institut f{\"u}r Physik, Technische Universit{\"a}t Dortmund,
  D-44221
  Dortmund, Germany\\
  $^{3}$ Department of Mathematical Sciences, University of Liverpool,
  Liverpool, L69 3BX, United Kingdom\\
  $^{4}$ TH Division, CERN, Geneva 23, Switzerland \\
  $^{5}$ LAPTH, Universit\'e de Savoie, CNRS, F-74941 Annecy-le-Vieux,  France\\
  $^{6}$ LPTHE, Universit\'e Paris 6, CNRS, F-75252 Paris, France }

\title{Flavour Violation in Anomaly Mediated Supersymmetry Breaking}

 \abstract{We study squark flavour violation in the
  anomaly mediated supersymmetry broken (\amsb{}) minimal
  supersymmetric standard model. Analytical expressions for the
  three-generational squark mass matrices are derived. We show that
  the anomaly-induced soft breaking terms have a decreasing amount of
  squark flavour violation when running from the GUT to the weak
  scale.  Taking into account inter-generational squark mixing, we
  work out non-trivial constraints from $\Bsg$ and $B_s \to \mu \mu$,
  which complement each other, as well as $B \to \tau \nu$ decays.  We
  further identify a region of parameter space where the anomalous
  magnetic moment of the muon and the $\Bsg$ branching ratio are
  simultaneously accommodated.  Since anomaly mediation is of the
  minimal flavour-violating type, the generic flavour predictions for
  this class of models apply, including a \ckm-induced (and hence
  small) $B_s -\bar B_s$-mixing phase.  }

\preprint{DAMTP-2008-112\\LTH 811\\DO-TH 08/08\\LAPTH-1292/08\\
} 

\begin{document}

\section{Introduction}
\vspace{-2mm}

Increasing precision in the calculation of sparticle effects is an
important part of theoretical preparation for the LHC\@. Much of this
work has concentrated on the \msugra{} scenario, where it is assumed
that the unification of gauge couplings at high energies is
accompanied by a corresponding unification in both the soft
\sy-breaking $\phi^*\phi$ scalar masses and the gaugino masses: and
also that the cubic scalar $\phi^3$ interactions are of the same form
as the Yukawa couplings and related to them by a common constant of
proportionality, the $A$ parameter.

Anomaly mediation (\amsb)~\cite{amsbbulk}--\!\cite{amsbthree} as the
main source of \sy\ breaking is an attractive alternative to the
\msugra{} paradigm. In \amsb, the $\phi^*\phi$ masses, $\phi^3$
couplings and gaugino masses are all determined by the appropriate
power of the gravitino mass multiplied by perturbatively calculable
functions of the dimensionless couplings of the underlying
\sic\ theory.  Moreover these functions are renormalisation group
(\RG{}) invariant, and the \amsb{} predictions are thus ultraviolet
insensitive~\cite{Arkani-Hamed:2000xj}.  Unfortunately the theory in
its simplest form leads to tachyonic sleptons and thus fails to
accommodate the usual electroweak vacuum state. There are many
different successful approaches which fix this problem, however.

There have been a number of studies of the sparticle spectrum in the
\amsb{} context but these have generally been carried out in the
approximation whereby third-generation Yukawa couplings only are
retained.  In this paper we consider {\it flavour\/} physics in the
\amsb{} context; aspects of this were considered in \reference{jftm}
for the $b \to s \gamma$ process, but there has been considerable
progress both on the experimental and theoretical side since then. We
will also show how \amsb{} satisfies the requirements of the principle
of {\it Minimal Flavour Violation}
(\mfv)~\cite{Chivukula:1987py}--\!\cite{Colangelo:2008qp}.  Moreover,
we will show that specific to \amsb{} there is a natural suppression
of flavour changing neutral current (\fcnc{}) effects related to the
size of the top quark Yukawa coupling at the electroweak scale.

We consider in some detail the critical calculation of the $\Bsg$
branching ratio, taking into account inter-generational squark mixing.
We show that for positive Higgs mass term $\mu$ the dependence of
$\bratio$ on $\tan\beta$, the ratio of the two Higgs vacuum
expectation values, is positively dramatic, because the charged Higgs
mass has a minimum for large $\tan\beta$ in the class of \amsb{}
models we are treating. As a result, for $\mu>0$, $\Bsg$ constrains
$\tan\beta$ to be relatively low; we nevertheless show that within
\amsb{} models it is possible for the supersymmetric contribution to
account for the current discrepancy between theory and experiment for
the muon anomalous magnetic moment.
We further analyse leptonic $B_s \to \mu \mu$ and $B \to \tau \nu$
decays within \amsb{}. For $B_s \to \mu \mu$ we take into account the
full flavour structure of the squark sector and include both chargino
and gluino contributions.  Despite \amsb{} being \mfv{}, the gluino
contributions induced by inter-generational down-squark mixing turn
out to be significant. We show that current data on the leptonic modes
are beginning to probe the $\mu<0$ branch.  Once higher statistics
become available these decays could provide decisive constraints on
the parameter space.

The plan of the paper is as follows: In Section~\ref{sec:review} we
review \amsb{} and squark flavour violation in minimal supersymmetric
models.  We present in Section~\ref{sec:analytical} analytical results
for the \amsb{} soft terms with full generational structure, showing
thereby how \amsb{} fulfils the \mfv{} principle.  We also assess the
effect on the squark mass spectrum of a number of solutions to the
tachyonic slepton problem. In Section~\ref{sec:numerics} we give
numerical estimates for the size of the flavour-mixing entries of the
squark mass matrices.  We further evaluate the constraints from the $B
\to X_s \gamma$ decay, work out implications for leptonic $B$-decay
observables in \amsb{} and comment on the anomalous magnetic moment of
the muon. In Section~\ref{sec:conclusions} we conclude.  In
Appendix~\ref{sec:detail} we provide details on the numerical
computation of the squark flavour-mixing parameters.

\section{Generalities \label{sec:review}}

We review \amsb{} in Section~\ref{sec:AMSB} and general squark flavour
violation within the \mssm{} in Section~\ref{sec:flavour}.

\subsection{Anomaly mediated supersymmetry breaking \label{sec:AMSB} }

For completeness and to establish notation, let us recapitulate some
standard results for the general case.  We take an ${\cal N}=1$
supersymmetric gauge theory with gauge group $\Pi_{\alpha} G_{\alpha}$
and with superpotential
\be
W(\Phi)=\frak{1}{6}Y^{ijk}\Phi_i\Phi_j\Phi_k+
\frak{1}{2}\mu^{ij}\Phi_i\Phi_j.
\label{eqf}
\ee
We also include the soft supersymmetry-breaking terms
\be
{\cal{L}}_{\rm soft}=-(m^2)^j{}_i\phi^{i}\phi_j-
\left(\frak{1}{6}h^{ijk}\phi_i\phi_j\phi_k+\frak{1}{2}b^{ij}\phi_i\phi_j
+ \frak{1}{2}M_{\alpha}\lambda_{\alpha}\lambda_{\alpha}+{\rm h.c.}\right) ,
\label{Aaf}
\ee
where we denote by $\phi$ the scalar component of the superfield
$\Phi$ and $\phi^i = (\phi_i)^*$. Here $M_{\alpha}$ are the gaugino
masses and $h$, $b$ and $m^2$ are the standard soft \sy-breaking
scalar terms.

The following set of results for soft \sy-breaking terms are
characteristic of \amsb{} and are \RG{} invariant~\cite{jjb}:
\besub
\bea
M_{\alpha} & = & \msoft \beta_{g_{\alpha}}/g_{{\alpha}},\label{eq:ama}\\
h^{ijk}& = & -\msoft\beta^{ijk}_{Y},\label{eq:amb}\\
(m^2)^i{}_j & = & \frac{1}{2}\msoft^2\mu\frac{d}{d\mu}\gamma^i{}_j,
\label{eq:amc}\\
b^{ij} & = & \kappa \msoft \mu^{ij} - \msoft \beta_{\mu}^{ij}.
\label{eq:amd}
\eea
\eesub
where $\gamma^i{}_j$ is the chiral superfield anomalous dimension
matrix, and $\beta_{g_{\alpha}}$, $\beta_Y$ are the $\beta$-functions
for the gauge and Yukawa couplings, respectively.  $\beta_Y$ is given
by
\be
\beta^{ijk}_{Y} = \gamma^i{}_l Y^{ljk} +  \gamma^j{}_l Y^{ilk} +
\gamma^k{}_l Y^{ijl}, 
\ee
and $\beta_{\mu}$ by a similar expression. At one loop we have
\besub
\bea
\label{beone}
\lf\beta_{g_{\alpha}}^{(1)} &=& 
g_{\alpha}^3\left[T(R_{\alpha})-3C(G_{\alpha})\right], \label{beone:1}\\
\lf\gamma^{(1)i}{}_j &=& 
\frak{1}{2}Y^{ikl}Y_{jkl}-2\sum_{\alpha}g_{\alpha}^2[C(R_{\alpha})]^i{}_j.
\label{beone:2}
\eea
\eesub
Here $R_{\alpha}$ is the group representation for $G_{\alpha}$ acting
on the chiral fields, $C(R_{\alpha})$ the corresponding quadratic
Casimir and $T(R_{\alpha}) = (r_{\alpha})^{-1}\Tr [C(R_{\alpha})]$ ,
$r_{\alpha}$ being the dimension of $G_{\alpha}$. For the adjoint
representation, $C(R_{\alpha})=C(G_{\alpha})I_{\alpha}$, where
$I_{\alpha}$ is the $r_{\alpha}\times r_{\alpha}$ unit matrix.
Obviously if the gauge group has an abelian factor, $G_1$ say, with
hypercharge matrix $\Ycal^i{}_j = \Ycal_i\delta^i{}_j$, then
$T(R_1)=\Tr[\Ycal^2]$, $\left[C(R_1)\right]^i{}_j=(\Ycal^2)^i{}_j$ and
$C(G_1) = 0$.

As we indicated in the introduction, \eqn{eq:amc}\ is unrealistic for
the sleptons; most phenomenology has been done by replacing it (at the
GUT scale) with
\be  (\mbar^2)^i{}_j =  \frac{1}{2}\msoft^2\mu\frac{d}{d\mu}\gamma^i{}_j
+m_0^2\delta^i{}_j , \label{eq:Amzero} 
\ee 
that is, by introducing a common scalar mass for the chiral super
multiplets.  We will call this model \mamsb\ in what follows.  There
have been a number of alternative approaches to the problem; for a
discussion see in particular~\reference{aprr}, and for the
phenomenology of {\it deflected\/} anomaly mediation see
\reference{rretal}.

One approach, first explored in detail in Ref.~\cite{jja}, and
subsequently by a number of
authors~\cite{Arkani-Hamed:2000xj}--\!\cite{Ibe:2004gh}, is to replace
\eqn{eq:amc}\ with
\be  (\mbar^2)^i{}_j = 
\frac{1}{2}\msoft^2\mu\frac{d}{d\mu}\gamma^i{}_j +\xi \Ycal_i\delta^i{}_j,
\label{eq:AUone} 
\ee 
where $\xi$ is a constant (with dimensions of $\hbox{mass}^2$) and
$\Ycal_i$ are charges corresponding to a $U(1)$ symmetry of the
theory.  The $\xi \Ycal$ term corresponds in form to a
Fayet-Iliopoulos (FI) $D$-term.  This alternative has the advantage
that it does not require us to postulate an independent source of
\sy\ breaking characterised by $m_0^2$; the new term in
\eqn{eq:AUone}\ can be derived in a natural way via the spontaneous
breaking of a $U(1)$ symmetry at high energies~\cite{amsbone,amsbtwo}.
 
For a discussion of how \eqn{eq:AUone}\ affects the \RG{} invariance
of the \amsb{}\ expressions see~\reference{amsbthree}. The outcome is
that if we work at a {\it specific\/} renormalisation scale (such as
$\msusy$) throughout, then we may use \eqn{eq:AUone}, with a specific
value of $\xi$, as long as the $U(1)$ represented by the charges
${\cal Y}$ has no mixed anomalies with the \sm{} gauge group.

An example of a way to provide a viable solution to this slepton
problem but retain \eqn{eq:amc}\ unaltered is to introduce $R$-parity
violating leptonic interactions, which provide positive sleptonic
$(\hbox{mass})^2$ contributions~\cite{Allanach:2000gu}.

Most applications of \amsb{} to the minimal supersymmetric standard
model (\mssm{}) and variants have employed \eqn{eq:ama},
(\ref{eq:amb}) and \eqn{eq:Amzero} or (\ref{eq:AUone}), and determined
the Higgs $B$ parameter (along with the $\mu$ term) by the
minimisation of the scalar potential. This reflects the fact that the
form of the $B$-term is more model dependent than the other soft
breaking terms; for a recent discussion see~\reference{amsbtwo}. In
fact \eqn{eq:amd}\ (with the arbitrary parameter $\kappa$) is the most
general form consistent with \RG\ invariance of the \amsb{} form of
soft supersymmetry breaking.

The \mssm{} (with right-handed neutrino superfields $\bar \nu$) admits
two independent, generation-blind and anomaly-free $U(1)$ symmetries,
one of which is of course $U(1)_Y$; it is convenient for our purposes
to parameterise them with the lepton doublet and singlet charges.  The
possible charge assignments are shown in \tabby{anomfree}; we will
call the additional symmetry $U(1)'$ in what follows.  Note that in
the effective theory below the scale of the right-handed neutrino mass
$U(1)'$ has no mixed anomalies with the \sm{} gauge group.
\TABLE[t]{
\begin{tabular}{|c c c c c c|} \hline 
& & & & & \\ 
$Q$ & ${\bar U}$ & ${\bar D} $
& $H_1$ & $H_2$ & ${\bar \nu}$ \\ 
& & & & & \\ \hline
& & & & & \\ 
$-\frac{1}{3}L$ & $-e-\frac{2}{3}L$  & $e+\frac{4}{3}L$
& $-e-L$ & $e+L$ & $-2L-e$ \\ 
& & & & & \\ \hline
\end{tabular}
\caption{\label{anomfree}Anomaly-free $U(1)'$ charges for arbitrary
  lepton doublet and singlet charges $L$ and $e$ respectively.
  $U(1)_Y$ corresponds to $L=-1/2$ and $e=1$. $\bar \nu$ is a \sm{} gauge
  singlet.}}

\begin{table}
\begin{center}
\begin{tabular}{|c c c c c c |} \hline   
& & & & & \\ 
$Q, {\bar U}$ & ${\bar D}$ & $\bar \nu$
& $H_1$ & $H_2$ & $N$ \\
 & & & & & \\ \hline
& & & & & \\ 
$e$ & $L$  & $2e-L$
& $-2e$ & $-e-L$ & $L+3e$ \\ 
& & & & & \\ \hline
\end{tabular}
\caption{\label{anomgutfree}Anomaly-free $U(1)_{SU(5)}$
  charges for arbitrary lepton doublet and singlet charges ($L$ and
  $e$ respectively) compatible with $SU(5)\times U(1)$. $N$, $\bar
  \nu$ are \sm{} gauge singlets.}
\end{center}
\end{table}
Alternatively, by introducing an additional \sm\ gauge singlet $N$ per
generation, appropriately charged under the $U(1)$ symmetry, and
completing the two Higgs multiplets to a $5$ and a $\fivebar$ (per
generation) we can have a charge assignment that is compatible with
grand unification to $SU(5)\times U(1)$ (see \tabby{anomgutfree}).
When we assess this possibility we will assume that only one pair of
Higgs doublets (and no Higgs triplets) survive in the effective field
theory below unification. So this case differs from the $U(1)'$ case
in that the $U(1)_{SU(5)}$ is anomalous in the low-energy theory; this
will affect the discussion of the \RG{} invariance of the soft terms
in what follows.

\subsection{Flavour structure of the \mssm{} Lagrangian}
\label{sec:flavour}

The quark chiral superfields of the \mssm{} have the following
$G_{SM}=SU(3)_c\times SU(2)_L\times U(1)_Y$ quantum numbers in the
{\tt SLHA2}~\cite{SLHA2} conventions, which we adopt:
\begin{equation}
Q:\,(3,2,\frak{1}{6}),\quad
{\bar U}:\,({\bar 3},1,-\frak{2}{3}), \quad {\bar D}:({\bar 3},1,\frak{1}{3}),
\label{fields}
\end{equation}
and the superpotential of the \mssm{} is written as
\begin{eqnarray}
W_Q&=& \epsilon_{ab} \left[ Q_i^{b}\, (Y_D)_{ij} \, H_1^a\, {\bar D}_{j}
+L_i^{b}\, (Y_E)_{ij} \, H_1^a\, {\bar E}_{j}+
Q_i^{a}\, (Y_U)_{ij}\, H_2^b\, {\bar U}_{j} - \mu H_1^a H_2^b\right].
\label{superpot}
\end{eqnarray}
Throughout this section, we denote $SU(2)_L$ fundamental
representation indices by $a,b=1,2$ and the generation indices by
$i,j=1,2,3$.  $\epsilon_{ab}=\epsilon^{ab}$ is the totally
antisymmetric tensor, with $\epsilon_{12}=1$. The $SU(3)$ colour
indices are suppressed. All \mssm{} running parameters are in the
$\drbar$ scheme~\cite{Jack:1994rk}.  We now tabulate the notation of
the relevant soft supersymmetry (\susy) breaking parameters. The
squark trilinear scalar interaction potential is
\begin{equation}
V_3 = \epsilon_{ab}
\left[
\tilde{Q}_{i_L}^{b}\, (T_D)_{ij} \, \tilde{d}^*_{jR}\, H_1^a +
\tilde{Q}_{i_L}^{a}\, (T_U)_{ij} \, \tilde{u}^*_{jR}\, H_2^b  
\right] + {\rm h.c.},
\end{equation}
where fields with a tilde are the scalar components of the superfield
with the identical capital letter. Note that the electric charges of
${\tilde u}_R$, ${\tilde d}_R$ are +2/3 and -1/3 respectively.  The
squark bilinear \susy-breaking terms are contained in the potential
\begin{equation}
V_2 =
{\tilde{Q}^*}_{iLa}\, (m_{\tilde Q}^2)_{ij}\, \tilde{Q}_{jL}^{a} +
 \tilde{u}_{iR}\, (m_{\tilde u}^2)_{ij} \, {\tilde{u}^*}_{jR} +
\tilde{d}_{iR}\, (m_{\tilde d}^2)_{ij} \, {\tilde{d}^*}_{jR} .
\label{softmasses}
\end{equation}

Eqs.~(\ref{superpot})--(\ref{softmasses}) are in the basis of flavour
eigenstates.  To discuss flavour violation we need to work in the
so-called super-\ckm{} basis, where the quark mass matrices are diagonal
and the squarks are rotated parallel to their fermionic partners. We
choose the following convention for the Yukawa couplings and for the
\ckm{} matrix $V$:
\begin{equation} 
\label{Yukawa-convention}
Y_U=V^T \mbox{diag}(\lambda_u, \lambda_c, \lambda_t)  , \qquad
Y_D=\mbox{diag}(\lambda_d, \lambda_s, \lambda_b)  ,
\end{equation}
where $\lambda_q$ denote the Yukawa couplings of the quarks in the
mass eigenstate basis. Under this convention the down-type
$SU(2)_L$-doublet squarks and the singlets are already in the
super-\ckm{} basis, while the up-type doublets need to be rotated. We
define the $6\times6$ mass matrices for the up-type and down-type
squarks as
\begin{equation}
{\cal L}^{\rm mass}_{\tilde q} ~=~ 
- \Phi_u^{\dagger}\,
{\cal M}_{\tilde u}^2\, 
\Phi_u
- \Phi_d^{\dagger}\,
{\cal M}_{\tilde d}^2\, 
\Phi_d~,
\end{equation}
where $\Phi_u = (\tilde u_L,\tilde c_L, \tilde t_L, \tilde u_R,\tilde
c_R, \tilde t_R)^T$ and $\Phi_d = (\tilde d_L,\tilde s_L, \tilde b_L,
\tilde d_R,\tilde s_R, \tilde b_R)^T$. The mass matrices read
\begin{eqnarray}
{\cal M}_{\tilde u}^2 &=&  \left( \begin{array}{cc}
  m^2_{\tilde U_L} + m^2_{u} + D_{u\,LL} ~~~
  &   
  \frac{v_2}{\sqrt{2}}\, {\widehat T}_U^\dagger  - \mu\, m_u\, \cot\beta \\\\
  \frac{v_2}{\sqrt{2}}\, {\widehat T}_U  - \mu^*\, m_u\, \cot\beta ~~~ &
       {{m}^2_{\tilde u}}^T + m^2_{u} + D_{u\,RR} \\                 
 \end{array} \right) \,\, ,
\label{massup}  \\
\nonumber\\
{\cal M}_{\tilde d}^2 &=&  \left( \begin{array}{cc}
   m^2_{\tilde D_L} + m^2_{d} + D_{d\,LL}~~~    
& 
   \frac{v_1}{\sqrt{2}}\, {T}_D^*  - \mu\, m_d\, \tan\beta      \\\\
   \frac{v_1}{\sqrt{2}}\, {T}_D^T  - \mu^*\, m_d\, \tan\beta~~~  &
   {m^2_{\tilde d}}^T + m^2_{d} + D_{d\,RR}
             \\                 
 \end{array} \right) \,\, .
\label{massdown}  
\end{eqnarray}
In the equations above, $v_1$ and $v_2$ are the vacuum expectation
values (VEVs) of the two Higgs doublets (with $\tan\beta \equiv
v_2/v_1$ and $v \equiv \sqrt{v_1^2 +v_2^2} \approx 246$ GeV), the
matrices $m_q$ (with $q=u,d$) are the diagonal quark masses and
$D_{q\,LL,RR}$ are flavour-diagonal D-term contributions. Furthermore,
$ m^2_{\tilde D_L} \equiv {m_{\tilde Q}^2}$, and we introduced the
$3\times 3$ matrices
\begin{equation}
m^2_{\tilde U_L} \equiv V \,{m}_{\tilde Q}^2\, V^\dagger \,,~~~
{\widehat T_{U}} \equiv T_{U}^T\,V^\dagger \,,
\label{eq:that}
\end{equation}
accounting for the rotation of the up-type doublets to the super-\ckm{}
basis.

\section{\amsb{} Squark Flavour   \label{sec:analytical}}

We derive and analyse the exact one-loop \amsb{} squark soft terms
with the full three-generational structure in
Section~\ref{sec:boundary}.  We then go on to show how the soft terms
are in \mfv\ form in Section~\ref{sec:MFV}.  In
Section~\ref{sec:tachyonic} we discuss the implications of various
solutions to the tachyonic slepton mass problem for the squark sector.

\subsection{Fully flavoured squark mass boundary conditions 
\label{sec:boundary}}

The one-loop anomalous dimensions for the quark and Higgs chiral
superfields are easily derived from \eqn{beone:2}\ and are given by
\besub
\begin{eqnarray}
(16 \pi^2)\gamma_Q^T   &=& Y_U Y_U^\dag + Y_D Y_D^\dag 
- \left(\frak{1}{30}g_1^2 + \frak{3}{2} g_2^2 
+ \frak{8}{3} g_3^2\right).\mbf{1} \, ,
\label{gamA} \\
(16 \pi^2)\gamma_U   &=&   2Y_U^\dag Y_U -  \left(\frak{8}{15}g_1^2 
+\frak{8}{3} g_3^2 \right).\mbf{1} \, ,
\label{gamB} \\
(16 \pi^2)\gamma_D   &=&  2Y_D^\dag Y_D  - \left( \frak{2}{15}g_1^2 
+ \frak{8}{3} g_3^2 \right).\mbf{1} \, ,
\label{gamC} \\
(16 \pi^2)\gamma_{H_2}   &=&   3\Tr \left(Y_U^\dag Y_U\right) 
- \frak{3}{10}g_1^2 - \frak{3}{2}g_2^2  \, ,
\label{gamD} \\
(16 \pi^2)\gamma_{H_1}   &=&   3\Tr \left(Y_D^\dag Y_D\right)
+\Tr \left(Y_E^\dag Y_E\right) 
- \frak{3}{10}g_1^2 - \frak{3}{2}g_2^2 \, , 
\label{gamE} 
\end{eqnarray}
\eesub
where $\mbf{1}$ is the identity matrix in flavour space. The quark
Yukawa $\beta$ functions are
\be
\beta_{Y_U} = Y_U\gamma_U + (\gamma_Q^T + \gamma_{H_2})Y_U,\quad 
\beta_{Y_D} = Y_D\gamma_D + (\gamma_Q^T + \gamma_{H_1})Y_D\label{betyB} , \\
\ee
from which expressions we obtain using \eqn{eq:amc}\ the following
leading-order results:
\besub
\begin{eqnarray}
\frac{(16 \pi^2)^2  (m_{\tilde Q}^2)^T}{\msoft^2} &=&
\left( -\frak{11}{50} g_1^4 - \frak{3}{2} g_2^4 + 8 g_3^4 \right).\mbf{1}
+ (Y_U Y_U^\dag) \left(3 \mbox{Tr}(Y_U Y_U^\dag) - \frak{13}{15}g_1^2 - 3
g_2^2 - \frak{16}{3} g_3^2 \right) \nonumber \\
&+& (Y_D Y_D^\dag) \left(3 \mbox{Tr}(Y_D Y_D^\dag) +  \mbox{Tr}(Y_E
Y_E^\dag)- \frak{7}{15}g_1^2 - 3 
g_2^2 - \frak{16}{3} g_3^2 \right) \nonumber \\
&+& Y_U Y_U^\dag Y_D Y_D^\dag +
 Y_D Y_D^\dag Y_U Y_U^\dag  
+ 3 (Y_U Y_U^\dag)^2 + 3 (Y_D Y_D^\dag)^2  , \label{fvamsba} \\
\frac{(16 \pi^2)^2 m_{\tilde u}^2}{\msoft^2} &=&
\left( -\frak{88}{25} g_1^4 + 8 g_3^4 \right).\mbf{1}
+ (Y_U^\dag Y_U) \left(6 \mbox{Tr}(Y_U Y_U^\dag) - \frak{26}{15}g_1^2 - 6
g_2^2 - \frak{32}{3} g_3^2 \right) \nonumber \\
&+& 2 Y_U^\dag Y_D Y_D^\dag Y_U 
+ 6 (Y_U^\dag Y_U)^2   ,\label{fvamsbb} \\
\frac{(16 \pi^2)^2 m_{\tilde d}^2}{\msoft^2} &=&
\left( -\frak{22}{25} g_1^4 + 8 g_3^4 \right).\mbf{1} \nonumber \\
&+& (Y_D^\dag Y_D) \left(6 \mbox{Tr}(Y_D Y_D^\dag) + 2 \mbox{Tr} (Y_E Y_E^\dag)-
\frak{14}{15}g_1^2 - 6 
g_2^2 - \frak{32}{3} g_3^2 \right) \nonumber \\
&+& 2 Y_D^\dag Y_U Y_U^\dag Y_D 
+ 6 (Y_D^\dag Y_D)^2   ,\label{fvamsbc} \\
\frac{16 \pi^2 T_U}{\msoft} &=& -
\left[ \left(3 \mbox{Tr} (Y_U Y_U^\dag) 
- \frak{13}{15} g_1^2 - 3 g_2^2 - \frak{16}{3} g_3^2\right).\mbf{1}
+3 Y_U Y_U^\dag + Y_D Y_D^\dag 
\right] Y_U  , \label{fvamsbd} \\
\frac{16 \pi^2 T_D}{\msoft} &=& -
\left[ \left(3 \mbox{Tr} (Y_D Y_D^\dag) + \mbox{Tr}(Y_E Y_E^\dag) 
- \frak{7}{15} g_1^2 - 3 g_2^2 - \frak{16}{3} g_3^2\right).\mbf{1}
 \right. \nonumber \\
&+&\left.  Y_U Y_U^\dag + 3Y_D Y_D^\dag 
\right] Y_D. \label{fvamsbe}
\end{eqnarray}
\eesub
The results agree in the dominant third-family flavour-conserving
limit with the expressions in Ref.~\cite{Gherghetta:1999sw}. Note the
presence in \eqn{fvamsba}\ of a $ Y_U Y_U^\dag$ term. As remarked, for
instance, in~\reference{Paradisi:2008qh}, such a term can lead to
sizeable contributions to \fcnc{} phenomena, if its coefficient is of
${\cal O}(1)$. We will see presently, however, that squark flavour
mixing in \amsb{} is in fact naturally suppressed in the
low-$\tan\beta$ region.

From the exact one-loop formulae for the squark soft terms in
Eqs.~(\ref{fvamsba})--(\ref{fvamsbe}) we can derive relations
displaying the flavour structure and suppression from the \ckm{} matrix
elements $V_{ij}$ explicitly. In the approximation that we retain only
the third-generation Yukawa couplings we find
\begin{eqnarray}
\left( m_{\tilde Q}^2 \right)_{ij}
&=& \frac{\msoft^2}{(16 \pi^2)^2} \left[ \delta_{ij}
\left( -\frak{11}{50} g_1^4 - \frak{3}{2} g_2^4 + 8 g_3^4 \right)
+ V_{ti}^* V_{tj} \lambda_t^2 ( \hat \beta_{\lambda_t} -\lambda_b^2 )
\right. \nonumber \\ 
&+& \left. \delta_{i3} \delta_{j3} \lambda_b^2 
( \hat \beta_{\lambda_b}-\lambda_t^2) +\lambda_t^2 \lambda_b^2 
( \delta_{j3} V_{ti}^* V_{tb} +\delta_{i3} V_{tj} V_{tb}^*  ) \right] ,  
\label{fvamsba-latb} \\
&&\nonumber\\
\left( m_{\tilde u}^2 \right)_{ij} &=&
\frac{\msoft^2}{(16 \pi^2)^2} \left[\delta_{ij} \left(
-\frak{88}{25} g_1^4 + 8 g_3^4 \right)
+2\, \delta_{i3}\delta_{j3} \lambda_t^2 
\left( \hat \beta_{\lambda_t}-\lambda_b^2 (1-|V_{tb}|^2) \right) 
\right] , \label{fvamsbb-latb} \\
&&\nonumber\\
\left( m_{\tilde d}^2 \right)_{ij} &=&
  \frac{\msoft^2}{(16 \pi^2)^2} 
\left[ \delta_{ij} \left(-\frak{22}{25} g_1^4 + 8 g_3^4 \right)+
2 \,\delta_{i3} \delta_{j3} 
\lambda_b^2 \left( \hat \beta_{\lambda_b} -\lambda_t^2 (1- |V_{tb}|^2) 
\right) \right] \, ,\label{fvamsbc-latb} \\
&&\nonumber\\
\left( T_U \right)_{ij} &=& - \delta_{j3} \frac{\msoft}{16 \pi^2} 
\lambda_t \left[ V_{ti} (\hat \beta_{\lambda_t}  -\lambda_b^2) 
+   \lambda_b^2 \delta_{i3} V_{tb} \right]  ,
\label{fvamsbd-latb} \\
&&\nonumber\\
\left( T_D \right)_{ij} &=& 
- \delta_{j3} \frac{\msoft}{16 \pi^2} \lambda_b
\left[ \delta_{i3} (\hat \beta_{\lambda_b}-\lambda_t^2) 
+ \lambda_t^2 V_{ti} V_{tb}^* \right].
\label{fvamsbe-latb}
\end{eqnarray}
Here, $\hat \beta_{\lambda_t}$ and $\hat \beta_{\lambda_b}$ are
defined through the beta functions of the top, $\hat \beta_{\lambda_t}
\equiv 16 \pi^2 \beta_{\lambda_t}/\lambda_t$, and bottom, $\hat
\beta_{\lambda_b} \equiv 16 \pi^2 \beta_{\lambda_b}/\lambda_b$, Yukawa
couplings, respectively, with one-loop expression in our approximation
given as
\begin{eqnarray}
\hat \beta_{\lambda_t}& =& 6 \lambda_t^2 +\lambda_b^2 - C_t  \,, \\
\hat \beta_{\lambda_b}& =& 6 \lambda_b^2 +\lambda_{\tau}^2 +\lambda_t^2 - C_b\,,
\end{eqnarray}
where 
\besub
\bea
C_t &=& \frak{13}{15}g_1^2 + 3g_2^2 + \frak{16}{3} g_3^2\,,\\
C_b &=&  \frak{7}{15}g_1^2 + 3g_2^2 + \frak{16}{3} g_3^2\,.  
\label{eqCb} 
\eea
\eesub
Note that $\hat \beta_{\lambda_t}, \hat \beta_{\lambda_b} <0$ in the
physical region. Incidentally, we remark that, when the
renormalisation scale approaches $\mgut$, $(m_{\tilde u}^2)_{33}$
turns negative as $\hat \beta_{\lambda_t}$ in Eq.~(\ref{fvamsbb-latb})
becomes more strongly negative.

Finally, performing the rotation of the up-type squark doublets to the
super-\ckm{} basis we find
\begin{eqnarray}
\left( m_{\tilde U_L}^2 \right)_{ij}&=&
\frac{\msoft^2}{(16 \pi^2)^2} \left[ \delta_{ij}
\left( -\frak{11}{50} g_1^4 - \frak{3}{2} g_2^4 + 8 g_3^4 \right)
+  \delta_{i3} \delta_{j3} \lambda_t^2 ( \hat \beta_{\lambda_t} -\lambda_b^2 )
\right. \nonumber \\ 
&+& \left. V_{ib} V_{jb}^* \lambda_b^2 
( \hat \beta_{\lambda_b}-\lambda_t^2) +\lambda_t^2 \lambda_b^2 
( \delta_{i3} V_{jb}^* V_{tb} +\delta_{j3} V_{ib} V_{tb}^*) \right] ,
\label{mULCKM}\\
\left( \widehat{T}_U \right)_{ij}&=& - \delta_{i3} \frac{\msoft}{16 \pi^2} 
\lambda_t \left[  \delta_{j3}  (\hat \beta_{\lambda_t}  -\lambda_b^2) 
+  \lambda_b^2 V_{jb}^* V_{tb} \right]~.
\label{TuCKM}
\end{eqnarray}

It is apparent from Eqs.~(\ref{fvamsba-latb})--(\ref{fvamsbe-latb})
and Eqs.~(\ref{mULCKM})--(\ref{TuCKM}) that inter-generational squark
mixing is suppressed by the off-diagonal entries of the \ckm{} matrix,
and that 1--3 mixing is smaller by one power of the Cabibbo angle with
respect to 2--3 mixing.

Of particular interest is the low- to moderate-$\tan\beta$ region,
i.e.~$\lambda_b \ll \lambda_t$. We see at once that, in that case, all
flavour violation in Eqs.~(\ref{fvamsba-latb})--(\ref{TuCKM}) would be
proportional to $\hat \beta_{\lambda_t}$.  It is thus a remarkable
feature specific to the \amsb{} soft terms that squark flavour
violation vanishes (at least for values of $\tan \beta$ where we may
neglect $\lambda_b$) as $\hat \beta_{\lambda_t}\to 0$, to the extent
that Eqs.~(\ref{fvamsba-latb})--(\ref{TuCKM}) remain a good
approximation at $\msusy$ (as we shall discuss, whether or not this is
true depends on our resolution of the tachyonic slepton problem).
Moreover, the value of $\tan\beta$ for which $\hat \beta_{\lambda_t}$
vanishes is close to the infrared quasi-fixed point (\irqfp) for
$\lambda_t$.  If we neglect the electroweak gauge couplings, the
\irqfp~\cite{Hill:1980sq}\ can be easily determined in the one-loop
approximation; it corresponds to
\be 
\frac{\lambda_t^2 (m_t)}{g_3^2(m_t)} = 
\frac{7}{18}\left(1
- \left(\frac{g_3^2 (M_X)}{g_3^2 (m_t)}\right)^{\frac{7}{9}}\right)^{-1},
\ee
$M_X$ being the scale of a Landau pole in $\lambda_t$.  For $M_X \sim
10^{16}$ GeV, of the order of the gauge unification scale, and
including electroweak corrections, we find that the \irqfp{} occurs at
$\lambda_t(m_t) \approx 1.1$, while $\hat \beta_{\lambda_t}$ vanishes
for $\lambda_t(m_t) \approx 1.2$.  Through $m_t =\lambda_t \,v\, \sin
\beta/\sqrt{2}$, we could predict $\tan\beta$ by inserting the
empirically measured top mass.  However, the resulting value of $\tan
\beta$ is very sensitive to higher-order corrections, therefore we
refrain from doing so here.  We instead estimate that for $1.0 \lae
\lambda_t(M_Z) \lae 1.2$ we are somewhere in the region $1< \tan\beta
< 10$.

So we conclude that, at small to moderate $\tan\beta$, flavour mixing
in \amsb{} is quite naturally suppressed, and resides in the mass
matrix for the down-type squarks.

The \mfv{} flavour mixing implies that the first- and
second-generation squarks are highly degenerate.  Moreover, again
specialising to low $\tan\beta$, we see that the down squarks obey a
$3+2+1$ pattern, with three degenerate $SU(2)$-singlet squarks, two
degenerate doublet squarks and one $SU(2)$-doublet sbottom. The
down-squark left-right mixing vanishes in this approximation
($\lambda_b \to 0$).  The up-squark spectrum in \amsb{} is of the type
$2+2+1+1$: it contains the first-two-generation singlet and doublet
squarks, and two stops with left-right admixture.
 
The dominant third-family approximation in
Eqs.~(\ref{fvamsba-latb})--(\ref{TuCKM}) is accurate to the per-mill
level except in two cases: $(m_{{\tilde U}_L}^2)_{12}$ and
$(m_{{\tilde D}_L}^2)_{12}$. Of these, the former is off by a few tens
of percent, due to a significant contribution which, albeit suppressed
by $(\lambda_s/\lambda_b)^2$, is enhanced by four inverse powers of
the Cabibbo angle with respect to the contributions in \eqn{mULCKM}:
\begin{equation}
\left(\Delta m_{{\tilde U}_L}^2 \right)_{12} = \frac{\msoft^2}{(16 \pi^2)^2}\,
V_{us} V_{cs}^* \,\lambda_s^2 \,\left(6\lambda_s^2 +3 \lambda_b^2 
+ \lambda_\tau^2-C_b\right).
\end{equation}
On the other hand, $(m^2_{{\tilde D}_L})_{12}$ is accurate at the
few-percent level.

\subsection{\amsb{} and Minimal Flavour Violation \label{sec:MFV}}

The usual notion of \mfv{} is that the source of all flavour violation
stems from the Yukawa matrices. This principle can be implemented to
hold if the Lagrangian satisfies a global SU(3)$^5$ flavour symmetry
\cite{Chivukula:1987py}, under which the Yukawa matrices act as
spurions and transform non-trivially.  Consequently, if we assume
$R$-parity conservation, the \mssm{} soft scalar masses such as, e.g.,
the squark masses, can be written in powers of Yukawa matrices as
\cite{D'Ambrosio:2002ex}
\bea \label{eq:mq2g} 
({m}_{\tilde Q}^{2})^T &=& z^q_{1}\,\mbf{1}+z^q_{2}\,{Y}_{U}
{Y}_{U}^{\dagger}+z^q_{3}\,{Y}_{D}{Y}_{D}^{\dagger}+z^q_{4}\,
({Y}_{U}{Y}_{U}^{\dagger})^{2}+z^q_{5}\,({Y}_{D}
{Y}_{D}^{\dagger})^{2} \\
&+& (z^q_6\, {Y}_{D}{Y}_{D}^{\dagger}{Y}_{U}{Y}_{U}^{\dagger} + {\rm h.c})
+ \ldots\, \, ,\nonumber \\
{m}_{\tilde u}^{2} &=& z^u_{1}\,\mbf{1}+z^u_{2}\,
{Y}_{U}^{\dagger} {Y}_{U}+z^u_{3} \, Y_U^\dagger {Y}_{D} {Y}_{D}^\dagger Y_U
+z^u_{4} \, ({Y}_{U}^{\dagger}{Y}_{U})^{2}
+ \ldots\, \, ,
\label{eq:mu2g} \\
{m}_{\tilde d}^{2} &=& z^d_{1}\,\mbf{1}+z^d_{2}\,{Y}_{D}^{\dagger}{Y}_{D}
+z^d_{3}\, Y_D^\dagger {Y}_{U} {Y}_{U}^\dagger Y_D +z^d_{4}\,
({Y}_{D}^{\dagger}{Y}_{D})^{2}
+ \ldots\, \, ,
\label{eq:md2g} 
\eea
where the ellipsis stands for terms involving higher powers of the
Yukawa matrices.

By the use of Cayley-Hamilton identities, it has been shown in
Ref.~\cite{Colangelo:2008qp} that the expansion in Eq.~(\ref{eq:mq2g})
terminates after a finite number of terms. It is further argued that,
by appropriately fine-tuning the coefficients $z_i$, {\it any}
$3\times3$ hermitian matrix can be cast in the form of
Eq.~(\ref{eq:mq2g}). This means that {\em all} \/ the \mssm{}
parameter space could be considered as \mfv{} if one takes the spurion
definition \cite{D'Ambrosio:2002ex} at face value. Therefore, the
decompositions Eqs.~(\ref{eq:mq2g})--(\ref{eq:md2g}) themselves are not
restrictive unless we impose additional constraints, such as
controlled departure from flavour blindness,
\begin{equation}
\frac{|z_{i}^x|}{|z_1^x|} \lesssim  {\cal{O}}(1) \qquad \forall i \geq 2 
\,,~~{x=u,d,q} \, , \label{mfvdef}
\end{equation}
suppressing large hierarchies among the coefficients. 

From the one-loop results for the \amsb{} squark masses
Eqs.~(\ref{fvamsba})--(\ref{fvamsbc}) one can infer the \mfv{}
expansion parameters:
\begin{eqnarray}
z_1^q & =& \frac{m_{3/2}^2}{(16 \pi^2)^2} \left(  -\frak{11}{50} g_1^4 - \frak{3}{2} g_2^4 + 8 g_3^4 \right) , \label{amsbmfv1}\\
z_2^q &=&  \frac{m_{3/2}^2}{(16 \pi^2)^2} \left(3 \mbox{Tr}(Y_U Y_U^\dag) - \frak{13}{15}g_1^2 - 3
g_2^2 - \frak{16}{3} g_3^2 \right), \\
z_3^q &=&  \frac{m_{3/2}^2}{(16 \pi^2)^2} \left(3 \mbox{Tr}(Y_D Y_D^\dag) + \mbox{Tr}(Y_E
Y_E^\dag)- \frak{7}{15}g_1^2 - 3
g_2^2 - \frak{16}{3} g_3^2 \right) , \\
z_{4,5}^q & = & 3 z_6^q = 3  \frac{m_{3/2}^2}{(16 \pi^2)^2} ,  \\
 z_1^u & =& \frac{m_{3/2}^2}{(16 \pi^2)^2}
 \left(  -\frak{88}{25} g_1^4 + 8 g_3^4 \right) , \qquad
z_1^d  = \frac{m_{3/2}^2}{(16 \pi^2)^2}
 \left(  -\frak{22}{25} g_1^4 + 8 g_3^4 \right) , \\
 z_2^u & =& 2 z_2^q, \qquad
z_2^d  = 2 z_3^q, \\
z_4^u & = & 3 z_3^u =6 \frac{m_{3/2}^2}{(16 \pi^2)^2}, \qquad
 z_4^d  =  3 z_3^d =6 \frac{m_{3/2}^2}{(16 \pi^2)^2}, \label{mfvamsb2}
\end{eqnarray} 
%
where all other $z_i^{u,d,q}$
vanish.  Note that  $z_2^{u,d,q}$ and $z_3^q$ are negative, and that all
of the $z_i^{u,d,q}$ are real. Thus
 non-\ckm{} \cp-violating phases do not exist in this
sector in \amsb. One potential source for non-\ckm{} \cp-violating phases in
\amsb{} models is a phase associated with the Higgs $\mu$ and $B$
terms~\cite{jftm}; 
another is the  right-handed neutrino Yukawa matrix.

\FIGURE[t]{
\unitlength=1.1in
\epsfig{file=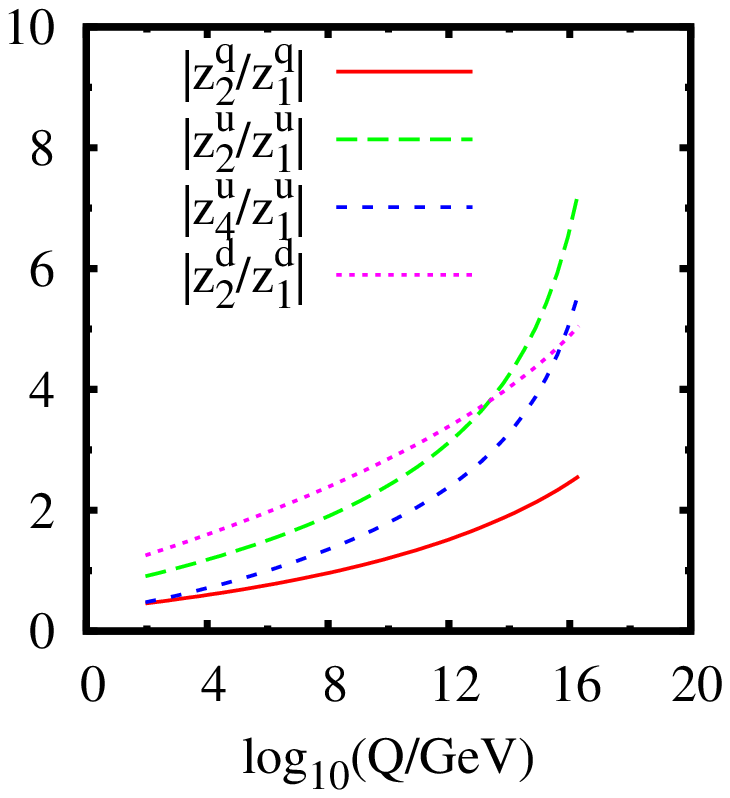, width=4.2in}
\caption{Selected \mfv{} ratios
  $|z_i^{u,d,q}/z_1^{u,d,q}|$ in pure \amsb{} for $\tan \beta=10$ as a
  function of the renormalisation scale $Q$.
 \label{fig:zplot}}}

In Fig.~\ref{fig:zplot} we plot some such selected ratios
$|z_i^{u,d,q}/z_1^{u,d,q}|$ as a function of the renormalisation scale
$Q$, which varies between $M_Z$ and $\mgut$.  We see an increase with
the scale in all the ratios, driven by the decrease of the
flavour-blind contributions proportional to $z_1^{u,d,q}$ towards the
GUT scale.  The suppression of flavour violation with decreasing scale
in the \mssm{} with general squark mixing has also been observed in
Refs.~\cite{meAndSteve,Paradisi:2008qh}.

The observed behaviour of the flavour coefficients $z_i^{u,d,q}$ in
\amsb{} is different from other common \mfv{} \mssm{} models.  While
\amsb{} is nowhere flavour blind (except for, at small $\tan\beta$, in
the limit ${\hat \beta}_{\lambda_t} \rightarrow 0$), both gauge
mediation and (by construction) \msugra{} have flavour-diagonal
sfermion masses at a certain high scale. In the latter models, the
$z^{u,d,q}_{i>1}$ parameters are induced by renormalisation group
evolution \cite{MandV}, and the ratios $|z_{i>1}^{u,d,q}/z_1^{u,d,q}|$
increase towards the weak scale. However, due to the automatic
suppression by loop factors (times logs) and the enhancement of the
$z_1^{u,d,q}$ terms by the gaugino contributions, the ratios
$|z_{i>1}^{u,d,q}/z_1^{u,d,q}|$ remain small, in agreement with
Eq.~(\ref{mfvdef}).

\subsection{Solutions to the tachyonic slepton problem \label{sec:tachyonic}}

An example of a scenario which fixes the tachyonic slepton problem
without disturbing Eqs.~(\ref{fvamsba})--(\ref{fvamsbe}) is provided
by~\reference{Allanach:2000gu}, where the \mssm{} is augmented by the
addition to the superpotential of (non \mfv) $R$-parity violating
couplings of the form $\lambda_{ijk}L_i L_j \bar E_k$.  These Yukawa
couplings provide positive contributions to the slepton squared masses
which can be sufficiently large, while leaving
Eqs.~(\ref{fvamsba})--(\ref{fvamsbe})\ unaffected at the scale of the
\susy-breaking terms, $\msusy$.  Other solutions to the tachyonic
slepton problem in which only the boundary conditions on the slepton
masses themselves are modified will generally affect the squark masses
as well, modifying their renormalisation group evolution below the
scale at which the additional slepton masses are switched on. However,
the slepton masses enter the one-loop $\beta$-functions for $m_{\tilde
  Q}^2$, $m_{\tilde u}^2$ and $m_{\tilde d}^2$ only via their
contribution to the $U(1)_{Y}$ Fayet-Iliopoulos (FI)
$S$-term~\cite{MandV} and consequently would have at most a small
effect on the running for these quantities.

On the other hand if we adopt the popular \mamsb\ solution of
\eqn{eq:Amzero} we must replace
\begin{equation}
m_{\tilde Q}^2 \to m_{\tilde Q}^2 +m_0^2\,\mbf{1}, \qquad
m_{\tilde u}^2 \to m_{\tilde u}^2  +m_0^2\,\mbf{1}, \qquad
m_{\tilde d}^2 \to m_{\tilde d}^2  +m_0^2\,\mbf{1}\label{zmsb}
\end{equation}
in Eqs.~(\ref{fvamsba})--(\ref{fvamsbc}) and apply the theoretical
boundary condition at the gauge unification scale $\mgut$. The \mssm{}
renormalisation group equations, which deviate from the pure \amsb{}
trajectory, must then be run down to the \susy{} scale $\msusy$ in
order to determine the mass spectrum. Note that even a
flavour-universal shift to the squark masses, such as the one in
Eq.~(\ref{zmsb}), affects the flavour-mixing mass parameters via the
running between $\mgut$ and $\msusy$. For instance, the beta function
for $(m^2_{\tilde{Q}_L})_{ij}$ (where $i \neq j$) contains a
piece~\cite{MandV} $(\beta_{m^2_{{\tilde Q}_L}})_{ij} = \sum_{l}
(m^2_{{\tilde Q}_L})_{ii} V_{il}^\dag ({\hat Y}_U)_{ll}^2 V_{lj} +
\ldots$, where ${\hat Y}_U$ is the diagonalised up-quark Yukawa
matrix.  Thus, a change to the flavour-universal piece of the squark
mass matrix $(m^2_{{\tilde Q}_L})_{ii}$ induces a change in
$(m^2_{{\tilde Q}_L})_{ij}$.

With the $U(1)$-based solution of \eqn{eq:AUone} we should really
distinguish the two alternatives we consider. With the $U(1)'$ model
(Table~\ref{anomfree}) we have
\begin{equation}
m_{\tilde Q}^2 \to m_{\tilde Q}^2 -\xi \,\frac{L}{3}.\mbf{1}, \ \
m_{\tilde u}^2 \to m_{\tilde u}^2 -\xi \left(e + \frac{2}{3}L\right).
\mbf{1}, \ \
m_{\tilde d}^2 \to m_{\tilde d}^2 +\xi \left(e +\frac{4}{3}L\right).
\mbf{1} \, .\label{fiamsb}
\end{equation}
In this case the non-FI contributions to the masses retain \RG{}
invariance, in the sense that applying \eqn{fiamsb}\ at $\mgut$ with a
given $(\xi e,\xi L)$ pair corresponds to the same physics as applying
the same equation at $\msusy$ with a different pair.  For example,
with $\msoft = 40~\TeV$ and $\tan\beta = 10$, and fixing for
simplicity $\xi = 1$ TeV$^2$ at both scales, the choice $(e,L) =
(0.25, 0)$ at $\mgut$ corresponds to $(e,L) \approx (0.06, 0.09)$ at
$\msusy$. The reason this does not correspond simply to a
renormalisation of $\xi$ is that, as well as such a renormalisation, a
FI term associated with $U(1)_Y$ is generated when we run down from
$\mgut$. This FI term can be absorbed into the existing one by
redefining $L$ and $e$. For a detailed discussion see
Section~\ref{sec:numerics} and in particular Eq.~(3.17)
of~\reference{amsbthree}. The allowed region in the $(e,L)$ plane has
been discussed in~\reference{amsbone}, see Fig.~1 of that reference.
With $\msoft=40~\TeV$ and $\xi = 1$ TeV$^2$, one needs $L \gae 0.03$
and $e \gae 0.04$ (at $\msusy$) to avoid negative square masses for
the slepton doublets and singlets, respectively, and it transpires one
also needs $L + e \lae 0.17$ in order that the Higgs potential gives
rise to the electroweak vacuum.  Thus, values of $(e,L)$ of ${\cal
  O}(0.1)$ are viable.

With the alternative of $U(1)_{SU(5)}$ from \tabby{anomgutfree} we
have
\begin{equation}
m_{\tilde Q}^2 \to m_{\tilde Q}^2 + \xi e.\mbf{1}, \qquad
m_{\tilde u}^2 \to m_{\tilde u}^2 + \xi e.\mbf{1}, \qquad
m_{\tilde d}^2 \to m_{\tilde d}^2 + \xi L.\mbf{1} \, , \label{fjamsb}
\end{equation}
but in this case the non-FI contributions to the masses are {\it
  not\/} \RG{} invariant because the low energy theory has
$U(1)_{SU(5)}$ anomalies, so we must again apply the theoretical
boundary condition at $\mgut$ and run the \mssm{} RGEs down to the
weak scale.  As discussed in~\reference{amsbthree}, there are lower
limits on $L$ and $e$ comparable to those found in the $U(1)'$ case,
but also a dramatic difference in that increasing $(e,L)$ with $L
\approx e$ does not lead to loss of the electroweak vacuum. The reason
for this is that in this case the FI contributions to the square
masses of both Higgses are negative.  Of course, increasing $(e,L)$
scales up the squark and slepton masses, $|m^2_{H_{1,2}}|$ and hence
the superpotential Higgs mass parameter $\mu$, thus increasing the
fine tuning known as the little hierarchy problem.

In all three cases we anticipate that, because of the flavour-blind
nature of the modification of the scalar masses, our expectation that
flavour violation will be suppressed at low $\tan\beta$ will turn out
to be true; it is clear, of course, that if we were to use a $U(1)$
with family-dependent charges in Eq.~(\ref{fiamsb})\ or
Eq.~(\ref{fjamsb})\ we would compromise the \mfv{} structure and
inevitably face \fcnc{} problems~\cite{jjc}.

\section{Predictions of Squark Flavour Violation \label{sec:numerics}}

In order to quantify \amsb{} predictions for flavour violation, we use
{\tt SOFTSUSY3.0}~\cite{softsusy}, which includes full three-family
flavour mixing. We consider the range $\msoft=40 - 140$ TeV, where the
lightest supersymmetric particle mass is $m_{\chi_1^0}\sim 130-520$
GeV and the gluino mass is $m_{\tilde g}=800-3100$ GeV, in the
interesting range for LHC \susy{} discovery~\cite{amsbLHC}. There are
no direct \susy-search constraints conflicting with $3 < \tan \beta <
42$ and 40 TeV $<\msoft<$ 140 TeV, therefore this is the range taken.
See Appendix~\ref{sec:detail} for further details on input parameters
and the calculation.

\subsection{Flavour-changing squark mass insertions in \amsb}

We now calculate the flavour-changing squark mass insertions in \amsb.
First, the Lagrangian parameters are transformed to the super-\ckm{}
basis described in Section~\ref{sec:flavour}, by rotating the one-loop
corrected squark mass matrices by the same mixing matrix required to
diagonalise the quark Yukawa matrices at $\msusy$. We may then define
the usual flavour-violating mass-insertion parameters $\dq$ from the
entries of the 6$\times$6 squark mass matrices ${\mathcal M}_{\tilde
  u}^2$ and ${\mathcal M}_{\tilde d}^2$ defined in Eqs.~(\ref{massup})
and (\ref{massdown})
\begin{eqnarray}
(\delta^{q}_{ij})_{LL} &=& \frac{({\mathcal M}_{\tilde q}^2)_{ij}}{\sqrt{
({\mathcal M}_{\tilde q}^2)_{ii} ({\mathcal M}_{\tilde q}^2)_{jj}}}, \quad
(\delta^{q}_{ij})_{RR} = \frac{({\mathcal M}_{\tilde q}^2)_{i+3\ j+3}}{\sqrt{
({\mathcal M}_{\tilde q}^2)_{i+3\ i+3} ({\mathcal M}_{\tilde q}^2)_{j+3\ j+3}}},
  \nonumber \\
(\delta^{q}_{ij})_{LR} &=& \frac{({\mathcal M}_{\tilde q}^2)_{i\ j+3}}{\sqrt{
({\mathcal M}_{\tilde q}^2)_{i i} ({\mathcal M}_{\tilde q}^2)_{j+3\ j+3}}},
\label{deltadef}
\end{eqnarray}
with $i,j \in \{1,2,3\}$ and $q = u,d$.  In this section we shall
compare the \amsb{} prediction of $\dq$ originating from
Eqs.~(\ref{fvamsba})--(\ref{fvamsbe}) with the empirical bounds from
Ref.~\cite{masiero}.
\FIGURE[t]{
\unitlength=1.1in
\begin{picture}(6,2.5)(0,0)
\put(0.2,0){\epsfig{file=deltau.eps,width=6.5cm}}
\put(3,0){\epsfig{file=deltad.eps, width=6.5cm}}
\end{picture} 
\vspace{-3mm}
\caption{Magnitudes of selected flavour-violating mass insertions $\dq$ 
    in \amsb{} as functions of $\tan\beta$. When two curves are visible
    for the same $\dq$, the upper curve is for $\msoft=40$ TeV and the
    lower curve is for $\msoft=140$ TeV. 
\label{fig:flavconsts}}}

In Fig.~\ref{fig:flavconsts}a we show the $\tan\beta$ dependence of
the absolute values of the flavour-violating up-squark mass insertions
$(\delta^{u}_{12})_{LL}\,,~{(\delta^{u}_{13})_{LL}}\,,~(\delta^{u}_{23})_{LL}$
and $(\delta^{u}_{23})_{LR}$ in the ``pure'' \amsb{} scenario, where
we assume that Eqs.~(\ref{fvamsba})--(\ref{fvamsbc}) are unaffected by
the mechanism that fixes the tachyonic slepton problem; while in
Fig.~\ref{fig:flavconsts}b we show the corresponding results for the
down-squark sector. The two curves for $(\delta^{u,d}_{23})_{LR}$
visible in each plot correspond to $\msoft=40$ GeV (upper curve) and
$\msoft=140$ GeV (lower curve), respectively. Indeed,
Eqs.~(\ref{massup}), (\ref{massdown}), (\ref{TuCKM}),
(\ref{fvamsbe-latb}) and (\ref{deltadef}) imply that
$(\delta^{u,d}_{ij})_{LR}$ are inversely proportional to $\msoft$ for
$i\neq j$, whence the significant, ${\mathcal O}(100\%)$ dependence
upon the \susy-breaking scale. On the other hand,
Eqs.~(\ref{fvamsba})--(\ref{fvamsbc}) combined with
Eq.~(\ref{deltadef}) imply that there is no dependence of
$(\delta^{u,d}_{ij})_{LL,RR}$ on $\msoft$ [aside from logarithmic
  corrections coming from scale dependence of the right-hand side of
  Eqs.~(\ref{fvamsba})--(\ref{fvamsbc})]. In
Figs.~\ref{fig:flavconsts}a and~\ref{fig:flavconsts}b the curves for
$(\delta^{u,d}_{ij})_{LL}$ that correspond to the two different values
of $\msoft$ are practically overlaid. We also see from the figures
that the mass insertions in the up-squark sector show a significant
dependence on $\tan \beta$, while the dependence in the down-squark
sector is much less pronounced. The reason for this is quite
simple. We can see from Eqs.~(\ref{fvamsba-latb}) and (\ref{mULCKM})
that the down-squark sector off-diagonal elements are more sensitive
to $\lambda_t$ and the up-squark off-diagonal elements are more
sensitive to $\lambda_b$; but as $\tan\beta$ increases from $5$ to
$40$, $\sin\beta$ (and hence $\lambda_t$) scarcely changes but
$\cos\beta$ (and hence $\lambda_b$) changes by a factor of 12.

In our solutions to the slepton mass problem, the additional
contributions in Eqs.~(\ref{zmsb}), (\ref{fiamsb}) and (\ref{fjamsb})
affect only the diagonal terms of the squark mass matrices at the
scale at which they are applied (i.e., $\msusy$ for the $U(1)'$
solution and $\mgut$ for \mamsb\ and $U(1)_{SU(5)}$). For a model such
as the $U(1)'$ solution in Eq.~(\ref{fiamsb}), which preserves the
\RG{} invariance of the expressions for squark soft \susy-breaking
terms, the change in the magnitudes of the $\dq$ parameters with
respect to the pure \amsb{} case can be directly estimated by the
effect of the slepton mass fix on the diagonal squark mass parameters.
Thus, denoting $x_{ij}^q \equiv \sqrt{ ({\mathcal M}_{\tilde
    q}^2)_{ii} ({\mathcal M}_{\tilde q}^2)_{jj}}$~,
\begin{equation}
\frac{\Delta |(\delta_{ij}^{q})_{XY}|}{|(\delta_{ij}^{q})_{XY}|} \approx -
\frac{\Delta x^q_{ij}}{x_{ij}^q}, \label{eq:ddel}
\end{equation}
with $q = u,d$. Interestingly, in the $U(1)$-inspired solutions the
shifts $\Delta x^q_{ij}$ enter the up and down singlet and doublet
squark masses in a non-universal way, hence the relative size of
$\delta^q_{LR}$ versus $\delta^q_{LL}$ can be modified at this level.
 
The experimental upper bounds upon the $\dq$ parameters depend upon
the squark masses and the ratio of the gluino mass to the squark
masses.  In order to obtain a rough estimate, we have fitted the
constraints in Ref.~\cite{masiero} with a parabola to determine the
dependence upon the gluino/squark mass ratio (whereas there is a
simple scaling relation with the squark mass itself).  We detail some
of the larger $\delta^q$ parameters in Appendix~\ref{sec:detail} for
four \amsb\ variants.  However, the bottom line is that all $\dq$ are
easily within their experimental bounds, regardless of which tachyonic
slepton fix is taken.  \amsb{} is far from being ruled out on the
basis of these naive empirical flavour constraints, the closest to the
bound being $(\delta_{13}^d)_{LL} \sim \mathcal{O} (10^{-3})$, which
has a bound of $(\delta_{13}^d)_{LL}<0.16$~\cite{masiero}.  However,
the mass insertions that mix the second and third generations can
affect the prediction of the branching ratios for rare $B$ decays such
as $\Bsg$ and $B_s \rightarrow \mu\mu$, by mediating the $b\rightarrow
s$ transition in loops involving squarks.  In
Sections~\ref{sec:bsgamma} and~\ref{sec:bsmumu} we will examine these
important physical observables, whose uncertainties have been vastly
reduced since Ref.~\cite{masiero}.

\subsection{AMSB prediction for the charged Higgs mass}

The Higgs sector of the \mssm\ (for a review see, e.g.,
Ref.~\cite{anatomy}) contains two \cp-even neutral scalars $h$ and $H$,
a \cp-odd neutral scalar $A$ and a charged scalar $H^\pm$. One of the
\cp-even scalars as well as $A$ and $H^\pm$ have couplings to the
down-type fermions that are enhanced by $\tan\beta$ with respect to
the couplings of the SM Higgs boson. Thus, even in \susy-breaking
scenarios such as \amsb\ in which the super particles are typically
rather heavy, there can be sizeable contributions to rare $B$ decays
from diagrams involving the non-standard Higgs bosons, if the latter
are light and $\tan\beta$ is large \cite{Hewett:1996ct}.

For moderate-to-large $\tan\beta$ the non-standard \cp-even scalar is
close in mass to the \cp-odd scalar, whose mass is determined by $m_A^2
= 2\,B\,/\sin2\beta$ at tree level. The masses of the \cp-odd and
charged scalars are in turn related at tree level by $m_{H^\pm}^2 =
m_{A}^2+m_W^2$. It is therefore useful to investigate the \amsb{}
prediction for the charged Higgs boson mass $m_{H^\pm}$, bearing in
mind that we determine the soft \susy-breaking Higgs mass parameter
$B$ by minimisation of the scalar potential. Inserting the pure
\amsb{} expressions \cite{Gherghetta:1999sw} for $m_{H_1}^2$ and
$m_{H_2}^2$ in the tree-level formula for $m^2_{H^\pm}$ (see
e.g.~Ref.~\cite{bpmz}), and neglecting contributions controlled by all
Yukawa couplings other than $\lambda_t$ and $\lambda_b$, we obtain, in
the large-$\tan \beta$ limit,
\begin{equation}
\frac{(16\pi^2)^2}{m_{3/2}^2}~m_{H^\pm}^2 ~\approx  ~
K \,-\,  \left(3\lambdabar_b^2\,C_b  - 36\,\lambdabar_b^4\right)\,\tan^2 \beta 
\,+\, 18\,\lambdabar_b^4\,\tan^4 \beta,  
\label{eq:chHiggsmass}
\end{equation}
where $K$ is positive and does not depend on $\tan\beta$ at tree
level, $C_b$ is defined in \eqn{eqCb} and ${\bar \lambda}_b \equiv
\lambda_b\,\cos\beta$. Since at tree level ${\bar \lambda}_b =
\sqrt{2}\,m_b/ v$, the coefficient of $\tan^2\beta$ is negative and
\eqn{eq:chHiggsmass} predicts a minimum for $m_{H^\pm}$ at a certain
value of $\tan \beta$.  However, for an accurate prediction of the
position of the minimum we must take into account the $\tan
\beta$-enhanced threshold corrections \cite{HRS} to the relation
between the bottom mass and the bottom Yukawa coupling, as well as the
radiative corrections to the tree-level formula for $m^2_{H^\pm}$.

\FIGURE[t]{
\unitlength=1.1in
\epsfig{file=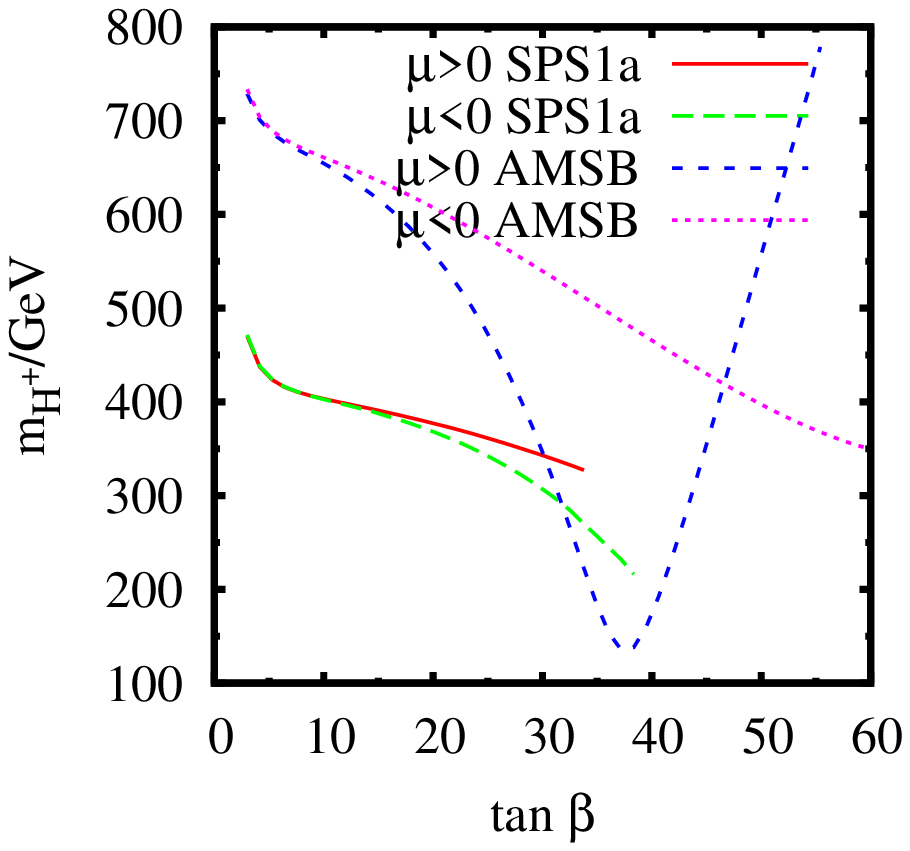, width=4.2in}
\caption{The charged Higgs boson mass as a function of $\tan\beta$ in 
\msugra{} model SPS1a and pure \amsb{} with $\msoft=40$ TeV and either 
sign of $\mu$. 
\label{fig:mhplus}}}

In Fig.~\ref{fig:mhplus} we show the full numerical dependence on
$\tan \beta$ of $m_{H^\pm}$, as computed by {\tt SOFTSUSY} for
``pure'' \amsb{} conditions, with $\msoft=40$ TeV and either sign of
$\mu$ (the $\mu>0$ curve terminates because the electroweak minimum of
the Higgs potential becomes unstable). The marked difference between
the curves corresponding to the two signs of $\mu$ is due to the fact
that the $\tan\beta$-enhanced threshold corrections, whose effect
depends on the sign of the product $\mgl\,\mu$, enhance ${\bar
  \lambda}_b$ for $\mu>0$ and suppress it for $\mu<0$. In the former
case the position of the minimum in $m_{H^\pm}$ is shifted towards
smaller values of $\tan\beta$, while in the latter we see no
stationary point up to $\tan\beta = 60$. For comparison we also show
$m_{H^\pm}$ as a function of $\tan\beta$ for the SPS1a \msugra{}
point~\cite{Allanach:2002nj}; the dependence on $\tan\beta$ is much
less marked. The curves end when the stau becomes tachyonic,
signalling an inappropriate scalar potential minimum.

\subsection{$\Bsg$ constraints }
\label{sec:bsgamma}
 
Flavour-changing neutral current processes are loop suppressed in the
\sm{} as well as in the \mssm. In the \sm{} the $\bsg$ transition is
mediated at one loop by diagrams involving $W$ boson and up-type
quarks. Additional one-loop contributions arise in the \mssm{} from
diagrams involving a charged Higgs boson and up-type quarks, a
chargino and up-type squarks and, in the presence of flavour violation
in the squark sector, a gluino and down-type squarks. The
contributions of diagrams with neutralinos and down-type squarks are
suppressed with respect to the gluino loops by the smaller gauge
coupling and by an accidental cancellation in the
magnetic-chromomagnetic mixing.

The current experimental value of the branching ratio for the $\Bsg$
decay is~\cite{hfag}
\begin{equation}
\bratio_{\rm exp}=(3.52\pm 0.23 \pm 0.09) \times 10^{-4}~,
\end{equation}
for a photon energy $E_\gamma>1.6$ GeV. The corresponding
next-to-next-to leading order (NNLO) \sm{} prediction that was published
two years ago reads \cite{supergroup}
\begin{equation}
\bratio_{\rm \,SM}=(3.15\pm 0.23) \times 10^{-4}~,
\end{equation}
and a recent update \cite{pgpg} of the calculation of the
normalisation factor for the branching ratio results in a modest
enhancement to $(3.28\pm 0.25) \times 10^{-4}$ (see also
Ref.~\cite{misiak}). In both cases, the error on the theoretical
prediction for the branching ratio is around 7\%. Inflating the
theoretical error to 10\% we accommodate -- rather conservatively --
for the additional uncertainty arising from the calculation of the
\susy{} contributions to the decay. Thus, at 95$\%$ C.L., we may
require $2.70 \times 10^{-4} < \bratio < 4.34 \times 10^{-4}$.

We use the public computer program {\tt SusyBSG 1.2}~\cite{susyBSG} to
obtain a next-to-leading order (NLO) prediction of $\bratio$ in the
\mssm. The program implements the results of Ref.~\cite{CDGG2} for the
two-loop gluon contributions to the Wilson coefficients of the
magnetic and chromomagnetic operators relevant to the $\bsg$
transition, and the full results of Ref.~\cite{DGS} for the two-loop
gluino contributions (accounting also for the $\tan\beta$-enhanced
charged-Higgs contributions first discussed in Ref.~\cite{bsg}). While
the two-loop contributions are computed in the approximation of
neglecting flavour mixing in the squark sector, the computation of the
one-loop contributions to the Wilson coefficients takes into account
the full flavour structure of the squark mass matrices. The relation
between the Wilson coefficients and $\bratio$ is computed at NLO along
the lines of Ref.~\cite{GM}, taking into account also the recent
results of Ref.~\cite{pgpg}. The free renormalisation scales of the
NLO calculation are adjusted in such a way as to mimic the NNLO
contributions that are not present in the calculation, reproducing the
central value of the \sm{} prediction of the branching ratio given in
Ref.~\cite{pgpg}.

\FIGURE[t]{
\unitlength=1.1in
\begin{picture}(6,2.5)(0,0)
\put(-0.8,0){\epsfig{file=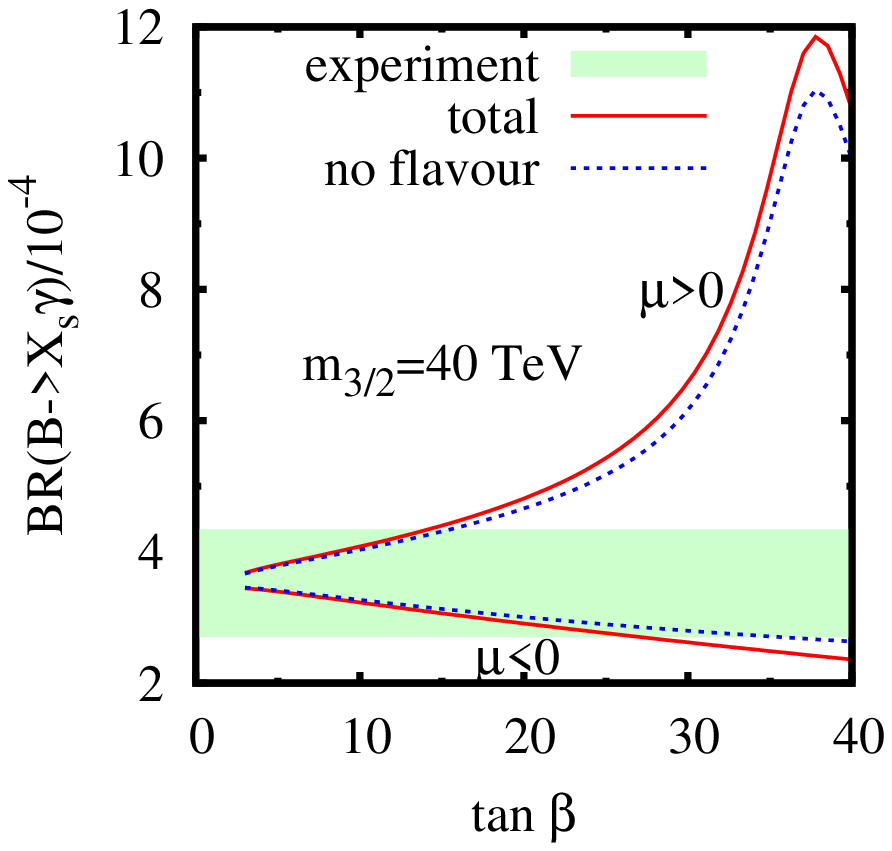,width=10cm}}
\put(1.6,0){\epsfig{file=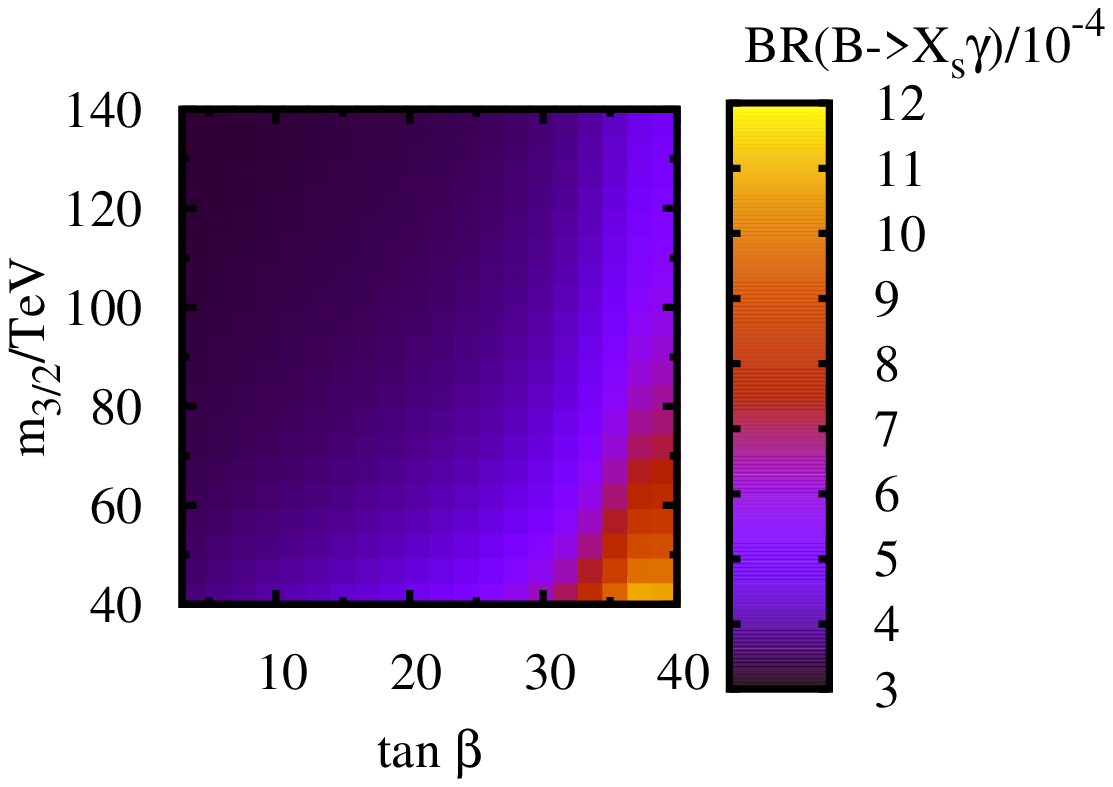, width=4.6in}}
\put(2.35,0.60){\epsfig{file=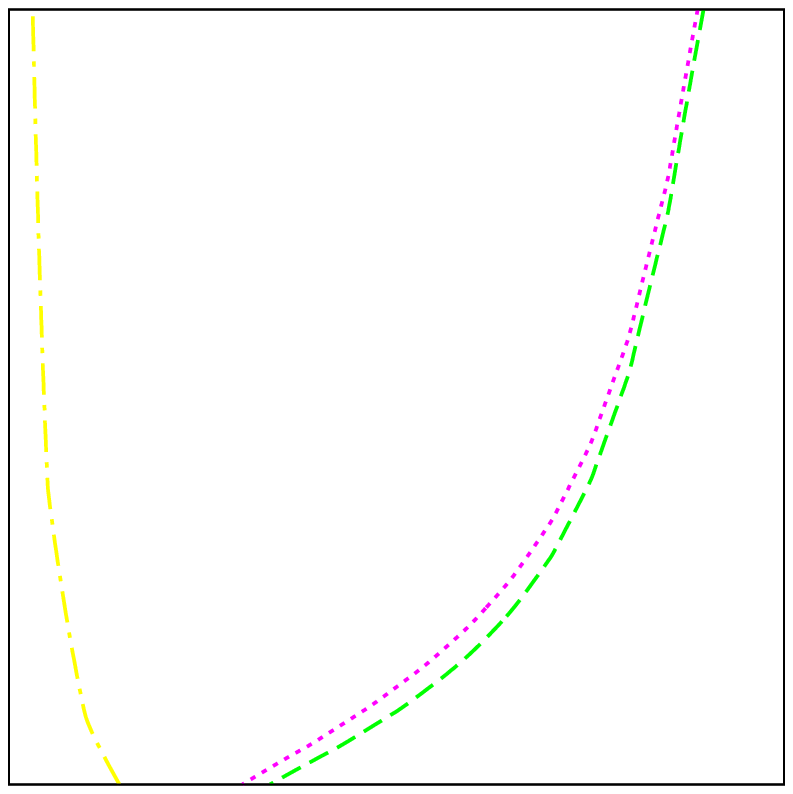, width=2.95in}}
\put(-0.3,2.3){(a)}
\put(2.2,2.3){(b)}
\end{picture}
\vspace*{-4.5mm}
\caption{Constraints on the \amsb{} parameter space from the rare
  decay $\Bsg$. (a) $\bratio$ as a function of $\tan \beta$ for
  $\msoft=40$ TeV and both signs of $\mu$. (b) $\bratio$ displayed as
  the background colour for $\mu>0$ in the $\tan \beta-\msoft$
  plane. For the explanation of the various curves in both panels see
  the text.
\label{fig:bsgScan}}}

Fig.~\ref{fig:bsgScan}a displays $\bratio$ as a function of $\tan
\beta$, for $\msoft=40$ TeV and either sign of $\mu$, assuming that
the squarks do not deviate from the pure \amsb{} trajectory. The red
(solid) curves include all effects in the calculation of the Wilson
coefficients, while the blue (dotted) curves ignore flavour-mixing
effects in the squark masses. The green shaded region represents the
95\% C.L.~limits on the branching ratio. The difference between the
curves corresponding to the two signs of $\mu$ is due to the
combination of two factors. First of all, as discussed above, the
$\tan\beta$-enhanced threshold corrections to the relation between the
bottom mass and the bottom Yukawa coupling result in a much lighter
charged Higgs boson -- thus an enhanced contribution to the Wilson
coefficients -- for $\mu>0$ (the peak in the branching ratio around
$\tan\beta \sim 37$ corresponds indeed to the minimum in $m_{H^\pm}$
shown in Fig.~\ref{fig:mhplus}). In addition, the contributions to the
Wilson coefficients from diagrams involving the top quark and the
charged Higgs boson and those from diagrams involving squarks and
charginos -- the latter depending on the sign of the product
$A_t\,\mu$, where $A_t \equiv (\widehat{T}_U)_{33}/\lambda_t$ --
interfere constructively for $\mu>0$ and destructively for $\mu<0$.
We remark that in the traditional \msugra\ scenario, in which $\mgl$
(and, in most cases, $A_t$) have opposite sign with respect to the
prediction of \amsb, the dependence of $\bratio$ on the sign of $\mu$
is reversed~\cite{bsg}.

The flavour-changing mass insertions $(\delta^d_{23})_{LL}$ and
$(\delta^d_{23})_{LR}$ mediate the $\bsg$ transition in one-loop
diagrams involving gluinos and down-type squarks. In addition,
$(\delta^u_{23})_{LL}$ can contribute a sizeable amount to one-loop
diagrams involving charginos and up-type squarks (the smallness of the
flavour-changing mass insertion being compensated by the fact that the
wino-strange-scharm vertex is not Cabibbo-suppressed).  From the
comparison between the red (solid) and blue (dotted) curves in
Fig.~\ref{fig:bsgScan}a we see that the flavour-violating effects have
a comparatively large effect (up to 10$\%$) on the predicted value of
$\bratio$ for large $\tan \beta$. We also see that, had we not
included squark flavour-violating effects in the calculation of
$\bratio$, we would have deduced that for $\mu>0$ the empirical limit
leads to $\tan \beta=15$, which is too weak by around 10$\%$.  For
$\mu<0$, neglecting squark flavour violation would have resulted on
the $\tan \beta$ bound being roughly 30\% too high.

Fig.~\ref{fig:bsgScan}b displays $\bratio$ as the background colour in
the $\tan \beta-\msoft$ plane, for $\mu>0$.  The yellow (dot-dashed)
contour on the left delimits the regions ruled out by the LEP2
Higgs-mass constraints\footnote{ 
LEP2 ruled out Standard Model Higgs masses of less than 114.4 GeV to
95$\%$ C.L.~\cite{lep2}. The same bound also applies, to a good
approximation, for the parameter space of \amsb{} investigated here. We
account for a 3 GeV theoretical error in the prediction of the Higgs
mass by plotting the bound for 111.4 GeV.}.
The red (dotted) contour on the right is the
bound on the $\tan \beta-\msoft$ plane obtained by applying the 95\%
C.L.~experimental upper bound on the branching ratio. The green
(dashed) rightmost contour is the bound that would be obtained if the
squark flavour mixing effects were ignored.  For a given value of
$\msoft$, the upper limit on $\bratio$ effectively provides an upper
bound on the parameter $\tan \beta$, because the \susy{} contribution
is enhanced for large $\tan \beta$.  We see that the strictest bound
is $\tan \beta < 13$ for $\msoft=40$ TeV but this relaxes to $\tan
\beta<35$ for $\msoft=140$ TeV, where heavier charged Higgs boson and
heavier sparticles provide a suppression of the \susy{} contribution
to $\bratio$.

\subsection{Implications for $B_s \to \mu \mu$ and future impact}
\label{sec:bsmumu}

\FIGURE[t]{
\epsfig{file=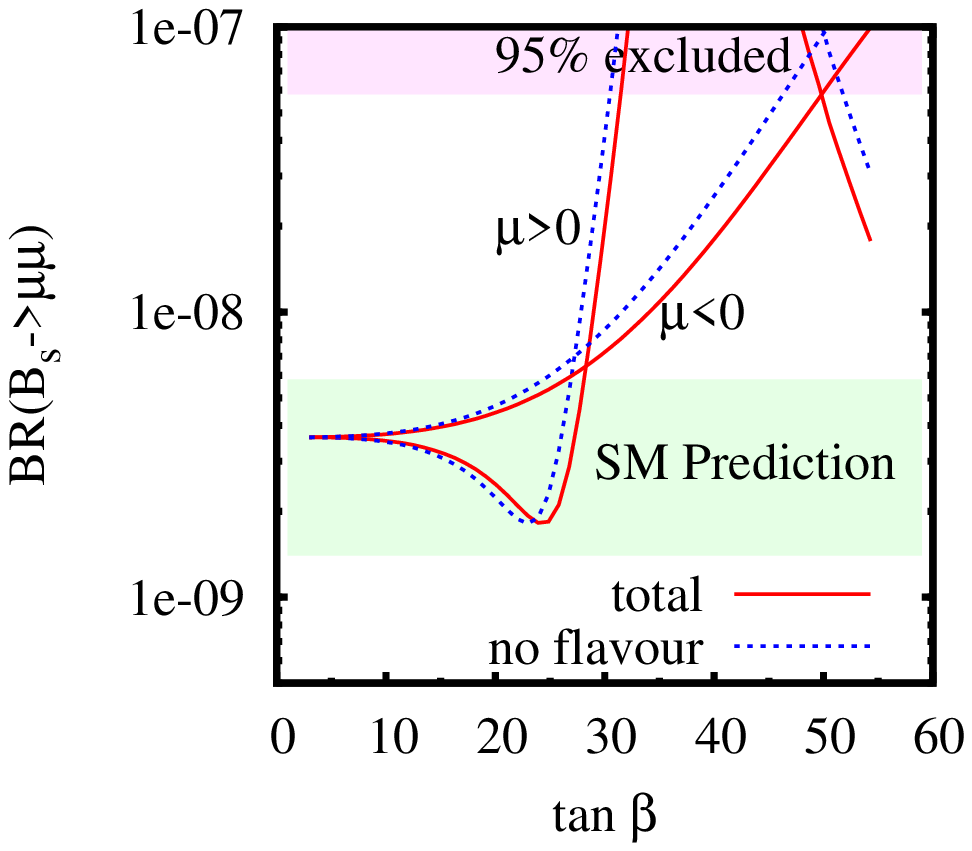,width=4.2in}
\caption{BR($B_s \rightarrow \mu\mu$) in pure \amsb{} with (``total'')
  or without (``no flavour'') squark flavour changing contributions to
  down-squark gluino loops for $m_{3/2}=40$ TeV and either sign of
  $\mu$.  Also shown is the \sm{} prediction and the current
  experimental upper bound \cite{prlcdf}. \label{fig:bmu1d} }}

The supersymmetric Higgs spectrum has a significant impact on the rare
leptonic decay $B_s \rightarrow \mu\mu$. Specifically, the decay
amplitude receives $\tan \beta$-enhanced contributions proportional to
$1/m_{A}^2$ from neutral-Higgs exchange \cite{bsmumu,Bobeth:2001sq}.
In our determination of the \mssm{} prediction for BR$(B_s \to \mu
\mu)$ we implemented the results of Ref.~\cite{Bobeth:2001sq} for the
subset of one-loop contributions involving up-type squarks and
charginos that are enhanced by $\tan^3\beta$, as well as the results
of Ref.~\cite{Bobeth:2002ch} for the one-loop contributions involving
down-type squarks and gluinos. The latter are relevant in the presence
of flavour mixing in the down squark sector; the dominant contribution
in \amsb{} stems from $(\delta^d_{23})_{LL}$, which, at $10^{-2}$, is
one of the largest mass insertions (see
Fig.~\ref{fig:flavconsts}). Finally, for the treatment of the $\tan
\beta$-enhanced, higher-order contributions that originate in the
corrections to the relation between the down-quark masses and Yukawa
couplings we followed Ref.\cite{roszkowski} (see also
Ref.~\cite{tanbetaresum}). We checked the relevant part of
our results against {\tt micrOMEGAS 2.1}~\cite{micromegas}, which
however does not include the effect of flavour mixing in the squark
sector.

Fig.~\ref{fig:bmu1d} shows BR$(B_s \to \mu \mu)$ in pure \amsb{} as a
function of $\tan\beta$, for $m_{3/2}=40$ TeV and either sign of
$\mu$. The red (solid) lines represent the total result, while the
blue (dotted) lines neglect the effect of flavour mixing in the squark
sector.  For the \sm{} branching ratio we obtain BR$(B_s \rightarrow
\mu\mu)_{\rm SM}=(3.6 \pm 0.9) \times 10^{-9}$, with the uncertainty
dominated by the one of the $B_s$-meson decay constant $f_{B_s}=0.24
\pm 0.03$ GeV \cite{Onogi:2006km}. For $\mu >0$ the effect of the dip
in $m_A$ (recall that $m_A \approx m_{H^\pm}$) around $\tan\beta \sim
35-40$ is clearly visible in the steep rise of the $B_s \rightarrow
\mu\mu$ branching ratio. For $\mu <0$ the $\tan\beta$-enhanced
corrections to the Higgs-quark-quark coupling cause a milder increase
with $\tan \beta$ (recall that in \amsb{} the relative sign between
$\mu$ and the gluino mass is opposite to the one in \msugra).
Our analysis also shows that -- contrary to what happens in $\bratio$
-- in BR$(B_s \rightarrow \mu\mu)$ the inclusion of squark flavour
mixing reduces the deviation from the \sm{} at large
$\tan\beta$. Here, the relative sign between the chargino and gluino
contributions is ${\rm sign}(A_t\,m_{\tilde
  g}\,(\delta^d_{23})_{LL})$, which is negative in \amsb{}.  The
effect of the gluino contribution is important and accounts for
changes up to a factor of two in the branching ratio.
We also show in Fig.~\ref{fig:bmu1d} the experimental 95\% C.L.~upper
bound BR$(B_s \rightarrow \mu\mu)<58 \times 10^{-9}$~\cite{prlcdf},
which is an order of magnitude above the \sm{} value.
Fig.~\ref{fig:bmu1d} shows that current $B_s\rightarrow \mu\mu$ data
is not as constraining as the $\Bsg$ branching ratio shown in
Fig.~\ref{fig:bsgScan}a, but if the experimental limit on
BR$(B_s\rightarrow \mu\mu)$ approaches the Standard Model prediction
in the future, for the $\mu<0$ branch, $B_s\rightarrow \mu\mu$ will
become more constraining than $\Bsg$.
\FIGURE[t]{
\unitlength=1.1in
\begin{picture}(6,2.5)(0,0)
\put(0.2,2.3){(a)}
\put(-0.6,0){\epsfig{file=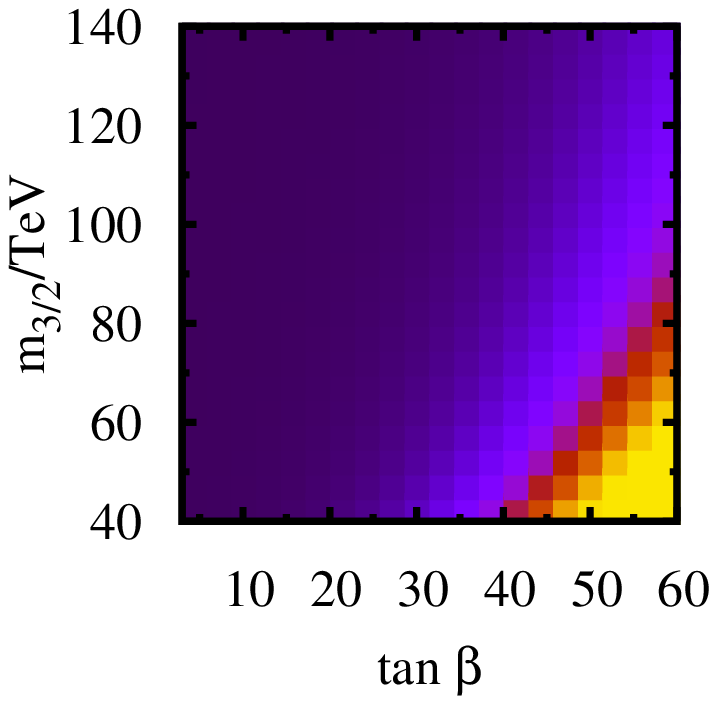, width=4.2in}}
\put(0.08,0.55){\epsfig{file=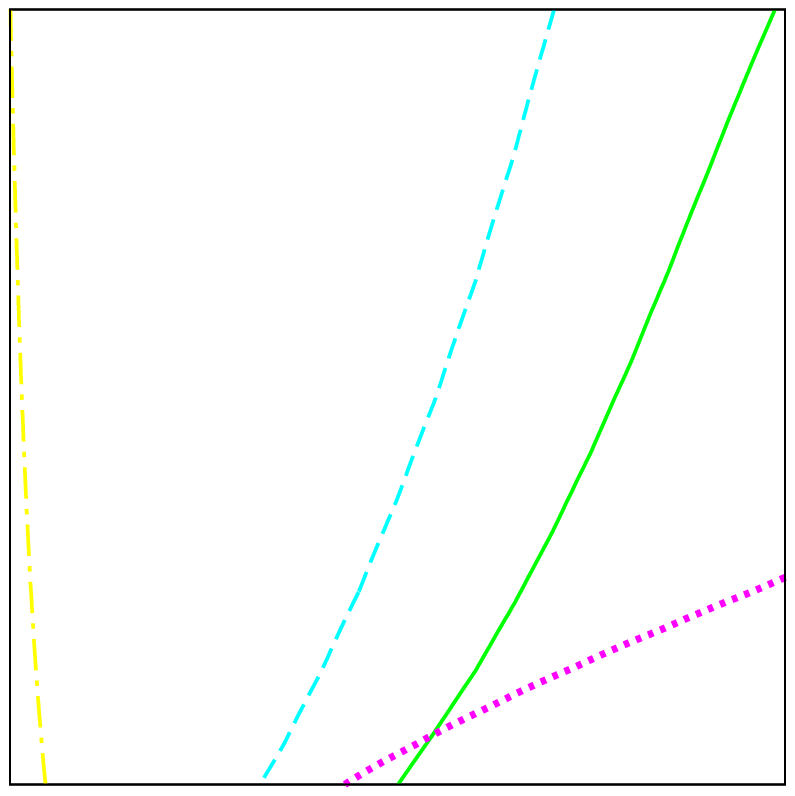, width=2.7in}}
\put(2.6,2.3){(b)} 
\put(1.8,0){\epsfig{file=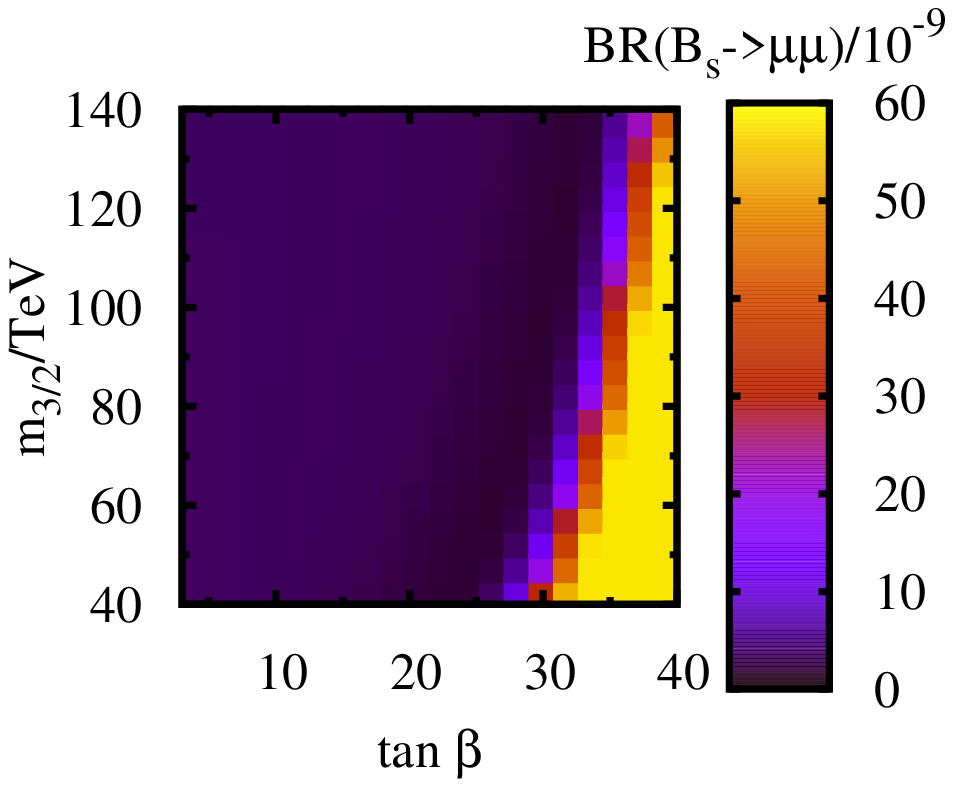, width=4.2in}}
\put(2.47,0.556){\epsfig{file=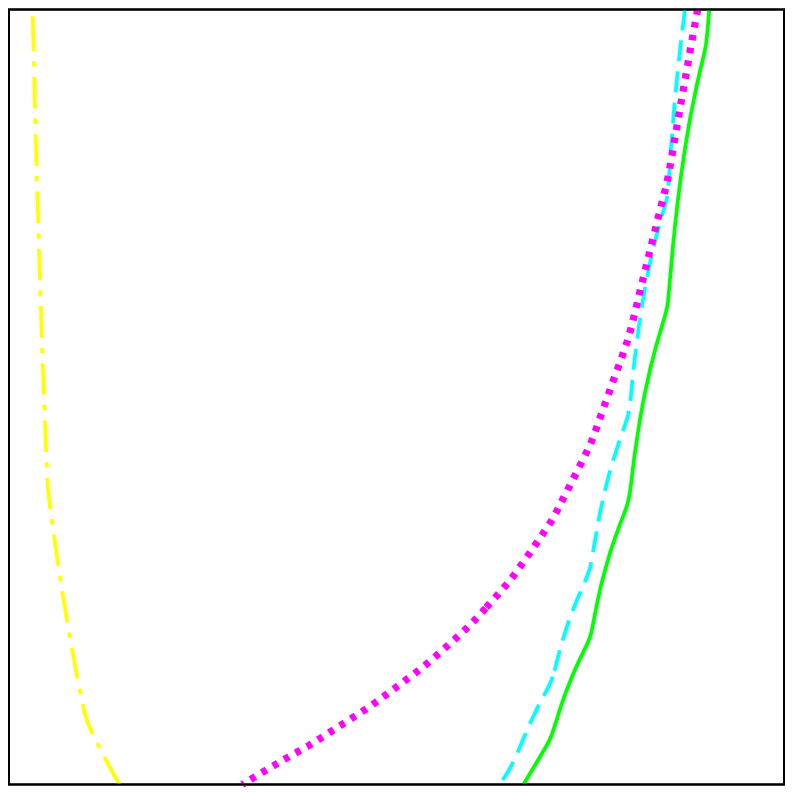, width=2.7in}}
\end{picture}
\vspace{-8mm}
\caption{BR($B_s \rightarrow \mu\mu$) in pure \amsb{} displayed as the
  background colour in the $\tan \beta-\msoft$ plane, for (a) $\mu<0$
  and (b) $\mu>0$.  For the explanation of the various curves in both
  panels see the text.
\label{fig:bmumu}}}

In Fig.~\ref{fig:bmumu} we show BR($B_s \rightarrow \mu\mu$) in pure
\amsb{} as the background colour in the $\tan \beta-m_{3/2}$ plane,
for (a) $\mu<0$ and (b) $\mu>0$. Constraints from a hypothetical
measurement of the branching ratio at $1 \times 10^{-8}$ (solid line)
and $5 \times 10^{-9}$ (dashed line) are given for
illustration. Superimposed on each panel are the boundaries of the
allowed region, which are as in Fig.~\ref{fig:bsgScan}b: 
the yellow
(dash-dotted) line delimits the parameter space allowed by the LEP2 Higgs
search, whereas
the magenta
(dotted) line marks the border of parameter space allowed by $\Bsg$.
Hence, for $\mu >0$, the $\Bsg$ constraint rules out the possibility
of a large $B_s \to \mu \mu$ enhancement at large $\tan \beta$.
Note that if we were to include also the
constraints on the muon anomalous magnetic moment, which requires a
positive $\mu$ term (see Section~\ref{sec:gm2}), we would predict the
$B_s \to \mu \mu$ branching ratio to not exceed its \sm{} value.

With improved data the rare leptonic mode will hence become
increasingly important. Searches for $B_s \to \mu \mu$ are ongoing at
the Tevatron collider and will commence at the LHC\@.  The LHCb
experiment will be able to exclude or discover new physics in $B _s
\to \mu \mu$ after one year, while ATLAS and CMS will be able to do so
after three years of operation \cite{Buchalla:2008jp}.

\subsection{Charged Higgs effects in $B \to \tau \nu$}
\label{sec:btaunu}

Substantial effects in the leptonic $B \to \tau \nu$ decays are
possible from charged Higgs exchange at large $\tan \beta$
\cite{Akeroyd:2003zr}.  It is customary to study the branching ratio
normalised to the \sm{} one, which yields a simple expression
\cite{Isidori:2006pk}
\begin{equation} \label{eq:Rtaunu}
R_{\tau \nu} \equiv \frac{\mbox{BR}(B \to \tau \nu)}
{\mbox{BR}(B \to \tau \nu)_{\rm SM}}
=\left( 1- \frac{m_B^2}{m_{H^\pm}^2} 
\frac{ \tan^2 \beta}{1+\epsilon_g  \tan \beta} \right)^2 .
\end{equation}
Here, $m_B$ denotes the mass of the $B$ meson and $\epsilon_g$ is the
gluino-induced correction to the relation between the mass of the
bottom quark and its Yukawa coupling.

In Fig.~\ref{fig:rbtaunu} we show $R_{\tau \nu}$ in \amsb{} for
$m_{3/2}=40$ TeV.  For $\mu >0$ the sharp peak around $\tan \beta \sim
37$ from the $m_{H^\pm}$ dip is clearly visible.  Using the stronger
constraint on $\tan\beta$ from $\bratio$, we predict $0.83 (0.96) <
R_{\tau \nu} \leq 1$ for $\mu >0 \, (\mu <0)$.  Thus, $R_{\tau \nu}$
is constrained to be below unity within \amsb{}, which is natural in
large-$\tan \beta$ \mfv{} scenarios \cite{Isidori:2006pk}.

\FIGURE[t]{
\unitlength=1.1in
\epsfig{file=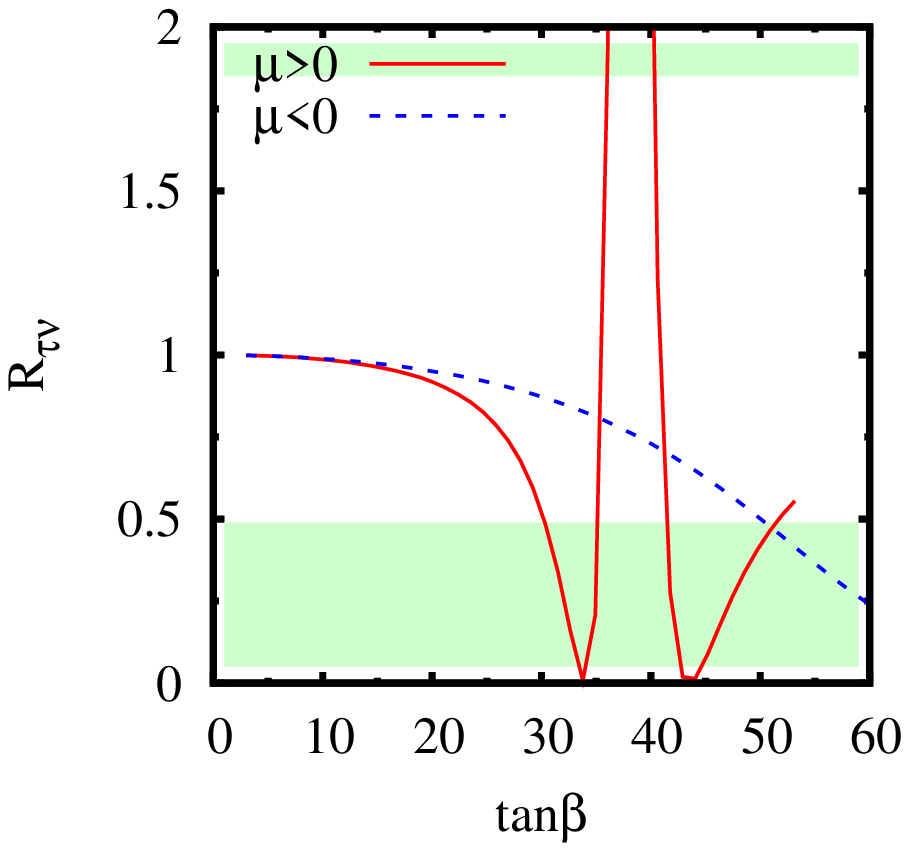, width=4.2in}
\caption{ The ratio $R_{\tau \nu}$ Eq.~(\ref{eq:Rtaunu}) in pure
  \amsb{} for $m_{3/2}=40$ TeV and either sign of $\mu$.  The green
  regions are disfavoured at the 2$\sigma$ level.
\label{fig:rbtaunu}} }

The branching ratio has been measured at the $B$-factories by Belle
and BaBar \cite{btaunuexp} $\mbox{BR}(B \to \tau \nu)=(1.51 \pm 0.33)
\times 10^{-4}$ with the average provided by \cite{hfag}.  With
$|V_{ub}|=(3.95 \pm 0.35) \times 10^{-3}$ \cite{PDG} and the $B$-meson
decay constant $f_B=0.216 \pm 0.022$ GeV \cite{Gray:2005ad} the \sm{}
prediction for the branching ratio is given as
\begin{equation}
\mbox{BR}(B \to \tau \nu)_{\rm SM} = 1.29 \times 10^{-4} 
\left( \frac{|V_{ub}|}{3.95 \cdot 10^{-3}}  \right)^2
\left( \frac{f_B}{0.216 \, \mbox{GeV}} \right)^2 ,
\end{equation}
with a net uncertainty of 19\%. For the ratio between experimental
result and SM prediction we obtain $R_{\tau \nu}^{\rm exp} =1.17 \pm
0.34$, where we added the uncertainties in quadrature.

We remark that the value of $|V_{ub}|$ used here results from
combining data on inclusive and exclusive $b \to u$ decays. Currently,
the individual determinations of $|V_{ub}|$ are not in perfect
agreement with each other, i.e., the exclusive modes prefer a lower
value than the inclusive ones.
Recent lattice computations \cite{Gamiz:2008iv} also give lower values
for $f_B$ and hence favour a somewhat larger $R_{\tau \nu}^{\rm exp}$
of 1.44$\pm$0.38, which is harder to accommodate within \susy.
Furthermore, the experimental situation for $B \to \tau \nu$ is also
still improving; at a high-luminosity $e^+ e^-$ machine \cite{superb},
a measurement of the branching ratio could perhaps be made with an
uncertainty of 10\% (for 10 $\mbox{ab}^{-1}$).
Given the situation, at present we cannot draw definite conclusions
for \amsb{} from $B \to \tau \nu$, but note that this mode has the
potential to become important in the future.

\subsection{Comment on $(g-2)_\mu$ \label{sec:gm2}}

In the \amsb{} context, having discussed $\bratio$, it behoves us to
comment on the \sic\ contribution to the muon anomalous magnetic
moment $\delta a_\mu$.  Relying on $e^+ e^-$ data for some of the hadronic
components, one finds~\cite{Miller:2007kk} that
\begin{equation}
\delta a_\mu \equiv \delta \frac{(g-2)_\mu}{2}  = (29.5 \pm 8.8) \times
10^{-10}
\end{equation}
is the discrepancy between the empirical value and the Standard Model
(\sm{}) prediction.  The one-loop gaugino contribution to this is given
at large $\tan\beta$ by~\cite{jftm,Moroi:1995yh,Marchetti:2008hw}
\be
a_{\mu}^{\susy} \approx \frac{m_{\mu}^2\mu\tan\beta}{16\pi^2}
\left(g_1^2 M_1 F_1 + g_2^2M_2 F_2 \right) ,
\label{eq:muong}
\ee
where $F_{1,2}$ are positive definite functions of the slepton,
chargino and neutralino masses, behaving like $1/\msusy^4$ in the
approximation that the relevant sparticles are degenerate in
mass. Thus for $M_1, M_2 > 0$, as is the case in \amsb, a
\sic\ explanation of the discrepancy between the \sm{} and experiment
favours $\mu > 0$. But we see from Fig.~\ref{fig:bsgScan} that
it is the $\mu > 0$ case that is restricted by $\bratio$.  So as
remarked, e.g., in~\reference{Martin:2001st}, this creates a potential
difficulty for explaining the discrepancy between theory and
experiment for $a_{\mu}$ using \amsb.
\FIGURE[t]{
\unitlength=1.1in
\epsfig{file=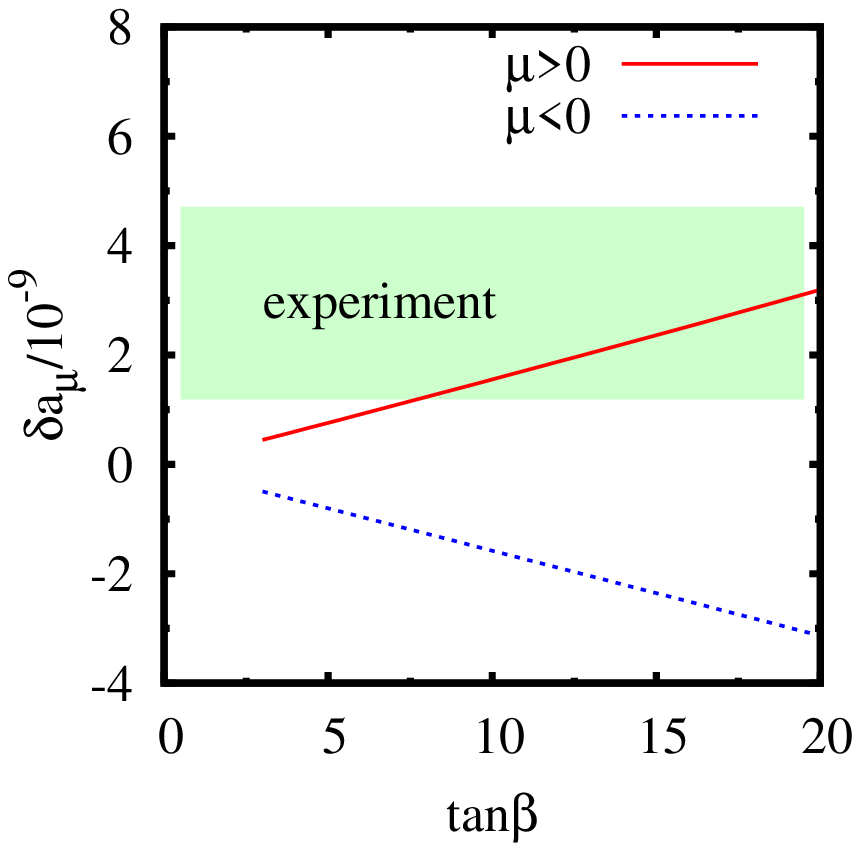, width=4.2in}
\vspace*{-0.5mm}
\caption{Supersymmetric contribution to the anomalous magnetic 
moment of the muon in $U(1)'$ 
\amsb, for $\msoft=40$ TeV and either sign of
$\mu$. The experimental constraint listed is at the 95$\%$ confidence
level. \label{fig:gm2}} }
Since $F_{1,2}$ depend upon the slepton masses, the prediction of
$\delta a_\mu$ in \amsb~models depends to a large extent upon the
slepton mass fix that is employed. In Fig.~\ref{fig:gm2}, we show such
a prediction for the $U(1)'$ fix. We take the one-loop results for
$\delta a_\mu$ from Ref.~\cite{Martin:2001st}, supplementing them with
the two-loop leading-log QED correction from
Ref.~\cite{Degrassi:1998es} and the $\tan \beta$-enhanced contribution
from Ref.~\cite{Marchetti:2008hw}. In the figure, it is clear how the
$\mu>0$ prediction in the red (solid) line fits the empirical 95$\%$
confidence level value of $\delta a_\mu$ for $\tan
\beta>8$. Comparison with Fig.~\ref{fig:bsgScan}a then shows a region
$8<\tan \beta<14$ which is compatible with both $\delta a_\mu$ and
$\Bsg$ constraints.

\vspace*{-2.5mm}
\section{Conclusions \label{sec:conclusions}}

We have investigated flavour violation in the squark sector in various
versions of \amsb; squark mixings are always readily calculable
because of the simple and constrained nature of supersymmetry breaking
terms in anomaly mediation.  The resulting supersymmetric
contributions to flavour-changing processes are \ckm-induced and hence
small.  The model thus is consistent with all observations of quark
flavour change.  Quark electric dipole moment constraints imply fairly
strict bounds on the imaginary phases on $|{\rm
  Im}(\delta^{u,d}_{11})_{LR}|<{\mathcal O}(10^{-6})$ ~\cite{masiero},
but these are easily satisfied due to the real coefficients
multiplying the Yukawa matrices in Eqs.~(\ref{fvamsbd}) and
(\ref{fvamsbe}).

At present, the branching ratio $\Bsg$ provides the most stringent
constraint on the model, and receives non-negligible supersymmetric
flavour corrections, affecting upper bounds on $\tan\beta$. As we
demonstrated, in the future, $B_s \to \mu \mu $ and $B \to \tau
\nu$ decays will provide complementary constraints.
We have also shown explicitly that there are regions of \amsb{}
parameter space that can accommodate the measurements of the $\Bsg$
branching ratio as well as the anomalous magnetic moment of the muon,
depending on the precise model for fixing the tachyonic slepton
problem. Indeed, a recent $\chi^2$ analysis of electroweak and baryon
precision observables favoured m\amsb{} over \msugra{} and minimal
gauge mediation~\cite{Heinemeyer:2008fb}; note, however, that this
analysis neglected inter-generational squark mixing effects.

Predictivity in the flavour sector makes the \amsb{} scenario an
attractive alternative to \msugra{}, whose family-universal pattern of
\susy-breaking sfermion masses is at best approximate.  It is not
immediately clear without further model building how the flavour
off-diagonal pieces of the sfermion mass squared matrices are
suppressed in order to give the \msugra{} pattern.  Moreover, \amsb{}
soft \susy-breaking terms are always present; the issue is whether, as
we have assumed here, they represent the dominant contributions to
\sy{} breaking.

Of course \amsb{} is not without its problems; the origin of the Higgs
$\mu$ term (and of the associated soft \susy-breaking $B$ term) is
model dependent, and in minimal versions the lightest supersymmetric
particle is the neutral wino, which represents a problematic dark
matter candidate. These difficulties are not insuperable, however (for
one approach see \reference{amsbtwo}). We believe that it is perhaps
time for \amsb{} to be afforded status comparable to \msugra{} in
modelling our expectations (or hopes) for what will be seen at the
LHC\@.  In any case, the two models should be easily discriminated in
the event of a supersymmetric signal at the LHC~\cite{amsbLHC} due to
their widely different predicted patterns of supersymmetric masses and
associated signals.

We close with some general remarks on quark flavour physics.  The
flavour changing signals of \amsb{} are \mfv{} in character: they
feature \ckm-induced \cp{} asymmetries, suppressed wrong-chirality
contributions and \ckm{} relations between $b \to s$ and $b \to d$
processes~\cite{Hiller:2003di}.  Because these models contain only a
minimal amount of flavour and \cp{} violation, their experimental
separation from the \sm{} background needs precise measurements,
feasible perhaps at super flavour factories
\cite{Buchalla:2008jp,superb}.

\section*{Acknowledgements} 
We thank C.~Bobeth, G.~Colangelo, G.~Isidori, A.~J\"uttner and
L.~Roszkowski for useful communication. This work has been partially
supported by STFC\@. BCA and DRTJ would like to thank the Aspen Center
of Physics for hospitality rendered during the conception and
commencement of this work. GH and DRTJ visited, and PS was based in
the CERN Theory Division during some of the subsequent
developments. BCA would like to thank the Technische Universit\"{a}t
Dortmund for hospitality offered and support under the Gambrinus
Fellowship while some of the work contained herein was performed.  The
work of GH is supported in part by the Bundesministerium f\"ur Bildung
und Forschung, Berlin-Bonn. The work of PS is supported in part by an
EU Marie-Curie Research Training Network under contract
MRTN-CT-2006-035505 and by ANR under contract BLAN07-2\_194882.

\appendix

\section{Numerical Detail of Squark Flavour Violation \label{sec:detail}}

In this appendix we collate the input parameters and detail of the
numerical calculation of the $\delta^q$ parameters as implemented in
{\tt SOFTSUSY 3.0}.  The sparticle pole masses receive one-loop
corrections to the flavour conserving pieces, and family mixing is
included at the tree level. {\tt SOFTSUSY} solves the \mssm{}
renormalisation group equations to two-loop order consistent with this
theoretical boundary condition and \sm{} data. Fermion masses and
gauge couplings are obtained at $M_Z$ using an effective field theory
of 3-loop QCD $\times$ 1-loop QED below $M_Z$. Our default \sm{} data
set contains the $\msbar$ quark masses $m_u(2 \mbox{~GeV})=2.4$ MeV,
$m_d(2 \mbox{~GeV})=4.75$ MeV, $m_s(2 \mbox{~GeV})=104$ MeV,
$m_c(m_c)=1.27$ GeV, $m_b(m_b)=4.23$ GeV~\cite{PDG}. The top quark
mass input is the pole mass, $m_t=172.4$ GeV~\cite{tevmtop}, and the
strong gauge coupling in the $\msbar$ scheme
$\alpha_s(M_Z)=0.1176$~\cite{PDG}.  We fix $M_Z=91.1876$ GeV to its
central value~\cite{PDG}, as well as the Fermi constant from muon
decays, $G_\mu=1.16637 \times 10^{-5}$ GeV$^{-2}$.
$\alpha(M_Z)=1/127.925$ is fixed to be the $\msbar$ value of the QED
gauge coupling. The \ckm{} mixing is parameterised by the Wolfenstein
parameters at their central empirical values~\cite{PDG}:
$\lambda=0.2258$, $A=0.814$, $\bar \eta=0.349$ and $\bar \rho=0.135$.

\TABLE[t]{
\begin{tabular}{|l | ccccc |}
\hline
~~~~~~~~~ & $U(1)'$ & pure & $\Re$pure & \mamsb\ & $U(1)_{SU(5)}$\ \\ \hline
$~~m_{{\tilde u}_L}$/GeV & 821& 817 & {817} & 853 & 877\\
$~~m_{{\tilde u}_R}$/GeV & 826& 822 & {822} & 857 & 881\\
$~~m_{{\tilde d}_L}$/GeV & 825& 820 & {820} & 856 & 880\\
$~~m_{{\tilde d}_R}$/GeV & 832 & 828 & {828} & 864 & 887 \\ 
$~~m_{{\tilde t}_L}$/GeV & 733& {729}  & {729} & 754 & 793\\
$~~m_{{\tilde t}_R}$/GeV & 636& {632}  & {632} & 645 & 703\\
$~~m_{{\tilde b}_L}$/GeV & 722& {718}  & {718} & 743 & 782\\
$~~m_{{\tilde b}_R}$/GeV & 821& 816 & {816} & 852 & 876\\ 
\hline
$(\delta^u_{13})_{LL}/10^{-5}$ & $-2.0+5.2i$& $-2.0+5.3i$ &$-5.6$ & $-5.7$ & $-5.1$\\
$(\delta^u_{23})_{LL}/10^{-4}$ & $-6.4+0.0i$ & $-6.6+0.0i$ &$-6.5$ & $-6.6$ &$-5.9$ \\
$(\delta^u_{23})_{LR}/10^{-5}$ & $-6.5-0.0i$& $-6.7+0.0i$ &$-6.7$ & $-6.4$ &$-5.8$ \\
$(\delta^d_{12})_{LL}/10^{-5}$ & $~~\,7.4+3.1i$&$~~\,7.4+3.1i$ & $~~\,5.4$ & $~~\,5.7$ & $~~\,4.8$\\
$(\delta^d_{13})_{LL}/10^{-3}$ &  $-2.0+0.8i$&$-2.0-0.8i$ & $-1.5$ & $-1.6$ & $-1.3$\\
$(\delta^d_{23})_{LL}/10^{-2}$ & $~1.0+0.0i$& $~~\,1.0+0.0i$ &$~~\,1.0$ & $~~\,1.0$ & $~~\,0.9$\\
$(\delta^d_{23})_{LR}/10^{-5}$ & $~~\,3.0+0.0i$& $~~\,3.0+0.0i$ &$~~\,3.0$ & $~~\,2.9$ & $~~\,2.7$\\
\hline
\end{tabular}
\vspace*{5mm}
\caption{Flavour-violating mass insertions $\dq$ for various different
  \amsb{} models for $\msoft=40$ TeV, $\mu>0$, $\tan \beta=10$.  The
  \mamsb\ point has $m_0=230$ GeV, whereas the $U(1)$ models both have
  $\xi=1$ TeV$^2$, with the $U(1)'$ model having $e=0.06$, $L=0.09$ at
  $\msusy$, and the $U(1)_{SU(5)}$ $e=L=0.1$ at $\mgut$. No loop
  corrections have been added to the masses.
  \label{tab:deltas}}}

In Table~\ref{tab:deltas} we display the full numerical determination
of the $\dq$ parameters for $\tan \beta=10$, $\mu>0$ and $\msoft=40$
TeV. Only $\dq$ parameters larger than $10^{-5}$ are listed. We
contrast the ``pure" \amsb{} prediction, where we assume that
Eqs.~(\ref{fvamsba})--(\ref{fvamsbc}) are unaffected by the mechanism
that fixes the tachyonic slepton problem (as is the case, e.g., for
the $R$-parity violating solution in Ref.~\cite{Allanach:2000gu}),
with models where the slepton mass problem has been fixed by other
means. In the model labelled \mamsb{} we introduce a common GUT-scale
scalar mass $m_0 = 230$ GeV as in \eqn{zmsb}. In the models labelled
$U(1)'$ and $U(1)_{SU(5)}$, with charges from Tables~\ref{anomfree}
and~\ref{anomgutfree}, respectively, the FI-term contributions to
\susy-breaking masses are added at $\msusy$ and $\mgut$, respectively,
setting $\xi=1~\TeV^2$, $e=0.06$ and $L=0.09$ in the first case and
$\xi=1~\TeV^2$, $e=L=0.1$ in the second.  In the upper section of the
table we display the square roots of the flavour-diagonal entries of
the squark mass matrices -- which, with a slight abuse of notation, we
denote as the masses of the corresponding squark species -- because
they shall be important for the following discussion. The
second-family squark masses are roughly degenerate with the
first-family squark masses of identical \sm{} quantum numbers.  For
the pure \amsb{} and the $U(1)'$ cases,
Eqs.~(\ref{fvamsba})-(\ref{fvamsbe}) may be applied directly at
$\msusy$ once the Yukawa and gauge couplings have been determined,
including complex phases in the definition of the \ckm{} matrix $V$. This
procedure neglects the scale dependence of $V$, but, between $M_Z$ and
$\msusy$, it is expected to be a small effect: $|\Delta
V_{ij}|/|V_{ij}| \leq {\mathcal O}(\lambda_t^2 \ln(M_Z/\msusy) /16
\pi^2)$.

For models which break the \RG{} invariance of the soft terms and have
boundary conditions imposed at $\mgut$ (here, the \mamsb{} and
$U(1)_{SU(5)}$ models), we use {\tt SOFTSUSY} to run all \mssm{}
parameters between $M_Z$ and $\mgut$. {\tt SOFTSUSY} does not
currently include complex phases in its RGEs, and when used in the
running-mode it fits $V$ to a real version with zero complex phase at
$M_Z$. The magnitude of each $V_{ij}$ is equivalent to the
corresponding fully complex $|V_{ij}|$ to better than the per-mille
level for all $V_{ij}$ except for $|V_{ts}|$, which is incorrect to
only 1$\%$, and $|V_{td}|$, which is incorrect by around 50$\%$
fractionally. Any $\dq$ parameters where the dominant contribution is
proportional to $V_{td}$ are therefore subject to this fractional
uncertainty. From Eqs.~(\ref{fvamsba-latb})-(\ref{TuCKM}), we see that
$(\delta^d_{12})_{LL}$ and $(\delta^d_{13})_{LL}$ are in this
category.

In order to investigate the size of inaccuracies due to the real
approximation, we employ the latter to calculate the pure \amsb{}
$\dq$ parameters, and list the results under the heading $\Re$pure in
Table~\ref{tab:deltas}.
The comparison between the `pure' and `$\Re$pure' approximations shows
that for all the $\delta^d$ parameters that involve the first
generation the discrepancy in absolute value between the exact and the
approximate results is of order 30\%--40\%.  For the $\delta^u$
parameters that mix the first and second generations the discrepancy
is of order 15\%--20\%. Finally, for the remaining $\delta^q$
parameters the real approximation reproduces the absolute value of the
complex result to better than 10\% accuracy. We expect that similar
uncertainties will be present in the \mamsb{} and $U(1)_{SU(5)}$ cases
on the results listed.

With our choice of parameters, the pure \amsb{} predictions for the
parameters $x_{ij}^q$ are of the order of 0.5 TeV$^2$, while the
additional contributions $\Delta x^q_{ij}$ are controlled by $(\xi
e,\xi L) = (0.06,0.09)$ TeV$^2$ (the smallish values of the charges
being necessary to ensure the correct breaking of the electroweak
symmetry). As a result, by comparing the second and third columns of
Table~\ref{tab:deltas} we see that the predictions for the $\dq$
parameters of the $U(1)'$ solution are rather close to those of the
pure \amsb{} solution: both the real and the imaginary parts of all
$\dq$ parameters agree to better than 10$\%$ fractional accuracy.

For solutions that break the \RG{} invariance of the soft \susy-breaking
terms, such as \mamsb{} and $U(1)_{SU(5)}$, the \RG{} evolution causes
the squark flavour-mixing parameters to depend on the form of the
tachyonic slepton fix. The \mamsb\ solution in Eq.~(\ref{zmsb}) makes
all squark mass-squared parameters larger by a common term $m_0^2$,
hence all $\dq$ smaller at the GUT scale where we assume this mass
contribution arises. However, $m_0^2 \approx 0.05$ TeV$^2$ does not
make a large difference to the squark masses for $\msoft=40$ TeV, as
the comparison between the \mamsb\ and $\Re$pure columns in
Table~\ref{tab:deltas} shows: the squark masses change by only a small
amount from their pure \amsb{} values (the largest being a $2\%$
fractional difference).  The above-mentioned RGE effects in squark
mixing parameters are evident for the \mamsb\ case, as some of the
small changes in the magnitudes of the $\dq$ parameters do not
correspond to a decrease as expected from squark mass effects alone.
However, the perturbation of the squarks away from their pure \amsb{}
trajectory, due to the addition of $m_0=230$ GeV to the scalar masses,
is small enough that Eqs.~(\ref{fvamsba})--(\ref{fvamsbe}) remain a
good approximation at the 10\% level.

Finally, the $U(1)_{SU(5)}$ solution in Eq.~(\ref{fjamsb}) allows for
larger values of the $(e,L)$ charges than the $U(1)'$ solution does,
without upsetting the breaking of the electroweak symmetry. Indeed, by
comparing the $\Re$pure and $U(1)_{SU(5)}$ columns in
Table~\ref{tab:deltas}, we see that with our choice $\xi e= \xi L=0.1$
TeV$^2$ (at $\mgut$) the deviations in the $\dq$ parameters from the
pure \amsb{} predictions are somewhat larger than in the other cases,
although still of the order of 10\%.

We see from Table~\ref{tab:deltas} that the other models in which the
slepton mass problem is fixed explicitly agree to roughly 10\%
fractional accuracy with the pure \amsb{} predictions for the $\dq$
parameters.  Had we raised our choice of $\xi$ from 1 TeV$^2$, or our
choice of $m_0$ from 230 GeV, we would start to see larger departures.
There is, however, clearly a non-negligible part of parameter space of
each model which reproduces the pure \amsb{} $\dq$ parameters and
which provides a solution to the tachyonic slepton problem.

\end{document}